\documentclass[12pt]{article}
\usepackage{epsfig}
\usepackage{amsmath}
\usepackage{hhline}
\usepackage{amssymb}
\usepackage{times}
\usepackage{cite}
\usepackage{setspace}

\usepackage{multirow}

\newlength{\dinwidth}
\newlength{\dinmargin}
\begin{document}  
\newcommand{\pom}{{I\!\!P}}
\newcommand{\reg}{{I\!\!R}}
\def\gsim{\,\lower.25ex\hbox{$\scriptstyle\sim$}\kern-1.30ex%
\raise 0.55ex\hbox{$\scriptstyle >$}\,}
\def\lsim{\,\lower.25ex\hbox{$\scriptstyle\sim$}\kern-1.30ex%
\raise 0.55ex\hbox{$\scriptstyle <$}\,}
\newcommand{\trm}{m_{\perp}}
\newcommand{\trp}{p_{\perp}}
\newcommand{\trmm}{m_{\perp}^2}
\newcommand{\trpp}{p_{\perp}^2}
\newcommand{\alps}{\alpha_s}
\newcommand{\sqrts}{$\sqrt{s}$}
\newcommand{\LO}{$O(\alpha_s^0)$}
\newcommand{\Oa}{$O(\alpha_s)$}
\newcommand{\Oaa}{$O(\alpha_s^2)$}
\newcommand{\Oo}{$O(1)$}
\newcommand{\pperp}{T}
\newcommand{\PT}{p_{\perp}}
\newcommand{\pt}{p_{_{T}}}
\newcommand{\JPSI}{J/\psi}
\newcommand{\PO}{I\!\!P}
\newcommand{\xbj}{x}
\newcommand{\xpom}{x_{\PO}}
\newcommand{\dgr}{^\circ}
\newcommand{\gev}{\,\mbox{GeV}}
\newcommand{\GeV}{\rm GeV}
\newcommand{\GeVS}{\rm GeV^2}
\newcommand{\gevsq}{\ensuremath{\mathrm{GeV}^2} }
\newcommand{\xp}{x_p}
\newcommand{\xpi}{x_\pi}
\newcommand{\xg}{x_\gamma}
\newcommand{\xgj}{x_\gamma^{jet}}
\newcommand{\xpj}{x_p^{jet}}
\newcommand{\xpij}{x_\pi^{jet}}
\renewcommand{\deg}{^\circ}
\newcommand{\qsq}{\ensuremath{Q^2} }
\newcommand{\et}{\ensuremath{E_t^*} }
\newcommand{\rap}{\ensuremath{\eta^*} }
\newcommand{\gp}{\ensuremath{\gamma^*}p }
\newcommand{\dsiget}{\ensuremath{{\rm d}\sigma_{ep}/{\rm d}E_t^*} }
\newcommand{\dsigrap}{\ensuremath{{\rm d}\sigma_{ep}/{\rm d}\eta^*} }
\newcommand{\pbarnt}{\,\mbox{{\rm pb$^{-1}$}}}
\newcommand{\Ptav}{\left\langle P_T \right\rangle}
\newcommand{\lt}{<}

\def\Journal#1#2#3#4{{#1} {\bf #2}, #4 (#3)}
\def\NCA{\em Nuovo Cimento}
\def\NIM{\em Nucl. Instrum. Methods}
\def\NIMA{{\em Nucl. Instrum. Methods} {\bf A}}
\def\NPB{{\em Nucl. Phys.}   {\bf B}}
\def\PLB{{\em Phys. Lett.}   {\bf B}}
\def\PRL{\em Phys. Rev. Lett.}
\def\PRD{{\em Phys. Rev.}    {\bf D}}
\def\PR{{\em Phys. Rev.}    }
\def\ZPC{{\em Z. Phys.}      {\bf C}}
\def\ZP{{\em Z. Phys.}      }
\def\EJC{{\em Eur. Phys. J.} {\bf C}}
\def\EJA{{\em Eur. Phys. J.} {\bf A}}
\def\CPC{\em Comp. Phys. Commun.}
\begin{titlepage}

\begin{flushleft}
{\tt DESY 09-162    \hfill    ISSN 0418-9833} \\
{\tt February 2010}                  \\
\end{flushleft}


\vspace*{50mm}

\begin{center}
\begin{Large}
{\boldmath \bf      
    Jet Production in $ep$ Collisions at Low $Q^2$ and\\ Determination of $\alpha_s$
}
\vspace*{10mm}

H1 Collaboration

\end{Large}
\end{center}

\vspace{2cm}

\begin{abstract}
\noindent
The production of jets is studied in deep-inelastic $e^{+}p$ 
scattering at low negative four momentum transfer squared 
$5<Q^2<100~\GeV^2$ 
and at inelasticity $0.2 < y < 0.7$
using data recorded by the H1 detector at HERA in the years 1999 and 2000, 
corresponding to an integrated luminosity of $43.5\pbarnt$. 
Inclusive jet, 2-jet and 3-jet cross sections as well as the
ratio of 3-jet to 2-jet cross sections are measured as a function of 
$Q^2$ and jet transverse momentum. 
The 2-jet cross section is also measured
as a function of the proton momentum fraction $\xi$.
The measurements are well described 
by perturbative quantum chromodynamics at next-to-leading order corrected for 
hadronisation effects and are subsequently used to extract the strong coupling 
$\alpha_s$.
\end{abstract}

\vspace{1.5cm}
\begin{center} Accepted by {\it Eur.Phys.J.C}\end{center}
\end{titlepage}

\newpage

\noindent
F.D.~Aaron$^{5,49}$,           
M.~Aldaya~Martin$^{11}$,       
C.~Alexa$^{5}$,                
V.~Andreev$^{25}$,             
B.~Antunovic$^{11}$,           
S.~Backovic$^{30}$,            
A.~Baghdasaryan$^{38}$,        
E.~Barrelet$^{29}$,            
W.~Bartel$^{11}$,              
K.~Begzsuren$^{35}$
% ...

\bigskip{\it
 \noindent
% ...
$ ^{5}$ National Institute for Physics and Nuclear Engineering (NIPNE) ,
          Bucharest, Romania \\
% ...
$ ^{11}$ DESY, Hamburg, Germany \\
% ...
$ ^{25}$ Lebedev Physical Institute, Moscow, Russia$^{ e}$ \\
% ...
$ ^{29}$ LPNHE, Universit\'{e}s Paris VI and VII, CNRS/IN2P3,
          Paris, France \\
 $ ^{30}$ Faculty of Science, University of Montenegro,
          Podgorica, Montenegro$^{ e}$ \\
% ...
$ ^{35}$ Institute of Physics and Technology of the Mongolian
          Academy of Sciences , Ulaanbaatar, Mongolia \\
% ...
$ ^{38}$ Yerevan Physics Institute, Yerevan, Armenia \\
% ...
$ ^{49}$ Also at Faculty of Physics, University of Bucharest,
          Bucharest, Romania \\
% ...
$ ^e$ Supported by the Deutsche Forschungsgemeinschaft \\
% ...
}

\newpage

\section{Introduction}

\begin{figure}[h]
\epsfig{file=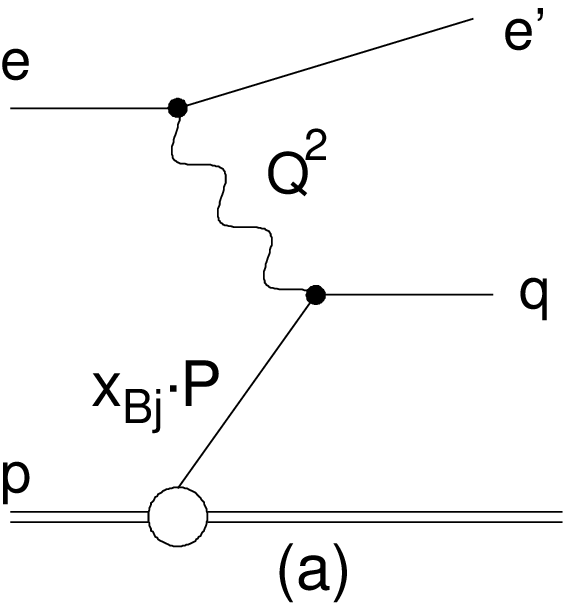,width=45mm}
\hspace*{8mm}\epsfig{file=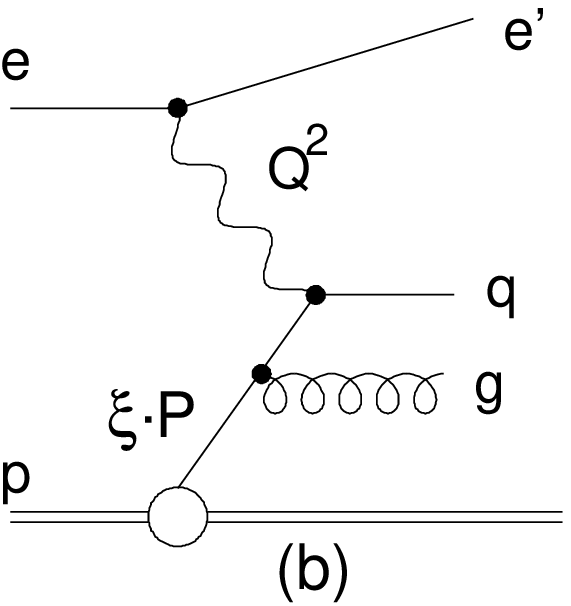,width=45mm}
\hspace*{8mm}\epsfig{file=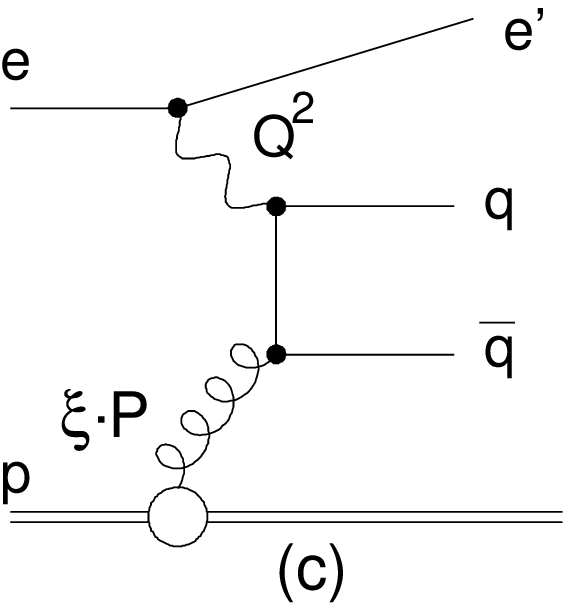,width=45mm}
\label{diagrams} 
\caption{\label{fig:feynborn} Deep-inelastic lepton-proton scattering at different orders in
$\alps$: (a) Born contribution  $\mathcal{O}(1)$
and  $\mathcal{O}(\alpha_s)$ processes (b) QCD Compton scattering and
(c) boson-gluon fusion.}
\end{figure}

Jet production in neutral current (NC) deep-inelastic scattering (DIS) 
at HERA provides an
important testing ground for Quantum Chromodynamics (QCD). 
While inclusive DIS gives only indirect information on the strong coupling via 
scaling violations of the proton structure functions, the production of jets allows 
a direct measurement of $\alpha_s$. The Born level contribution to 
DIS (Fig.~\ref{fig:feynborn}(a)) generates no transverse momentum in the Breit frame, 
where the virtual boson and the proton collide head on \cite{feynman}. 
Significant transverse momentum $P_T$ in the Breit frame is produced at leading order 
(LO) in the strong coupling $\alpha_s$ by the QCD Compton (Fig.~\ref{fig:feynborn}(b)) 
and boson-gluon fusion  (Fig.~\ref{fig:feynborn}(c)) processes. The latter dominates 
jet production for the range of the negative four momentum transfer squared 
of this analysis, $5<Q^2<100~\gevsq$, 
and provides direct sensitivity to the gluon density function of
the proton~\cite{Adloff:2000tq}.

Analyses of inclusive and multi-jet production in DIS were previously 
performed
at high 
$Q^2$ ($\gtrsim 100~\GeV^2$)
\cite{Adloff:2000tq, Aktas:2007pb, Aaron:2009vs, Chekanov:2006yc} and at low 
$Q^2$  ($\lesssim 100~\GeV^2$) \cite{Adloff:2000tq, Adloff:2002ew, Chekanov:2007dx} 
by the H1 and ZEUS Collaborations at HERA. 
In this paper new measurements of the inclusive jet, 
2-jet and 3-jet production cross sections, as well as the
ratio of 3-jet to 2-jet cross sections, are presented as a function 
of $Q^2$ and the jet transverse momenta in the
Breit frame, $P_T$, in the ranges $5<Q^2<100~\gevsq$ and $P_T>5~\GeV$.
The 2-jet cross section is also presented as a function of 
$\xi = x_{\rm Bj} (1+ M^2_{12} / Q^2 )$, which in LO
corresponds to 
the momentum fraction of the proton carried by the interacting parton  
(see Figs.\ref{diagrams}(b) and \ref{diagrams}(c)).
%
%
The variable $x_{\rm Bj}$ denotes the Bjorken scaling variable and $M_{12}$ the 
invariant mass of the two jets of highest $P_T$. The data correspond to higher 
integrated luminosity and a higher centre-of-mass energy than in the previous 
H1 analyses at low $Q^2$ \cite{Adloff:2000tq,Adloff:2002ew}. 
The larger data set together  with improved understanding of the hadronic 
energy measurement significantly  reduces the total uncertainty of the
cross section measurements. The results are compared
with perturbative QCD predictions at next-to-leading order (NLO) 
corrected for hadronisation effects, and $\alpha_s$ is extracted from a fit 
of the predictions to the data. 
These measurements allow the running of the strong coupling to be tested down to the
limits of the perturbative calculation. 
Together with the high $Q^2$ measurements~\cite{Aaron:2009vs}
the data test the running of $\alpha_s$ in the range of 
renormalisation scale $\mu_r$ between about $6$ and $70$~GeV.


\section{Experimental Method}
\label{section:experiment}

The data used for this analysis were recorded with H1 detector 
in the years 1999 and 2000, when
HERA collided positrons of energy 
$E_e = 27.6 \ {\rm GeV}$
with protons of energy 
$E_p = 920 \ {\rm GeV}$ giving a centre-of-mass energy
$\sqrt{s}=319~\rm GeV$.
%
The corresponding integrated luminosity is
$43.5$~pb$^{-1}$.

\subsection{The H1 detector}

A detailed description of the H1 detector can 
be found in \cite{Abt:1997hixv,Appuhn:1997na}. 
Here, a brief account of the components most 
relevant to the present analysis is given. 
The origin of the H1 coordinate system is the nominal
$ep$ interaction point. 
The direction of the proton beam defines the
positive $z$-axis (forward direction).
The polar angle 
$\theta$ is measured with respect to this direction. The pseudorapidity is 
defined as  $\eta=-\ln\tan(\theta/2)$.

In the central region  \mbox{($20^\circ\!<\!\theta\!<\!160^\circ$)}
the $ep$ interaction region is surrounded by a two-layered silicon strip detector~\cite{Pitzl:2000wz}
and two large concentric drift
chambers (CJCs), operated inside a $1.16$~T solenoidal magnetic field. 
The trajectories of charged particles are measured
in the central tracker with a transverse momentum resolution  of
$\sigma(p_T)/p_T=0.006\, p_T/$GeV$\, \oplus \,$0.02~\cite{Kleinwort:2006zz}.
Two additional drift chambers complement the CJCs by precisely 
measuring the $z$-coordinates of 
track segments and hence improve the determination of the polar angle. 
The central tracking detectors also provide triggering
information based on track segments measured in the $r$-$\phi$ plane of the
central jet chambers and on the $z$ position of the event vertex obtained from the double
layers of two multiwire proportional chambers (MWPCs).
The forward tracking detector and the 
backward drift chamber (BDC) measure
tracks of charged particles at smaller \mbox{($7^\circ\!<\!\theta\!<\!25^\circ$)} 
and larger \mbox{($155^\circ\!<\!\theta\!<\!175^\circ$)} polar angle
than the central tracker, respectively.

A finely segmented electromagnetic
 and hadronic liquid argon (LAr) calorimeter~\cite{Andrieu:1993kh} 
surrounds the tracking chambers. It has a
polar angle coverage of
\mbox{$4^\circ\!<\!\theta\!<\!154^\circ$} and full azimuthal acceptance.
The energy resolution is
$\sigma(E)/E=0.12/\sqrt{E/\gev}\oplus 0.01$
 for electromagnetic showers and 
$\sigma(E)/E=0.5/\sqrt{E/\gev}\oplus 0.02$
for hadrons, as
measured in test beams~\cite{Andrieu:1993tz}.
A lead-scintillating fibre spaghetti calorimeter (SpaCal)~\cite{Appuhn:1997na}
covers the backward region \mbox{$153^\circ\!<\!\theta\!<\!178^\circ$}.
Its main purpose is the detection of scattered positrons.
The energy resolution of the SpaCal for positrons is
$\sigma(E)/E=0.071/\sqrt{E/\gev}\oplus 0.01$.

 
The luminosity is measured via the Bethe-Heitler Bremsstrahlung
process $ep \rightarrow ep \gamma$, the final state photon being detected in a
crystal calorimeter at $z=-103$~m.

\subsection{Event and jet selection }
\label{sec:jetsel}

The data sample of this analysis was collected using a 
combination of triggers which 
require the scattered positron to be measured in the SpaCal, at least one 
high transverse momentum track ($p_\pperp>800\ {\rm MeV}$) to be 
reconstructed in the central tracking chambers and an event vertex to be 
identified by the MWPCs. The trigger efficiency is close to $100\%$
for the whole analysis phase space as determined from the 
data using independent triggers as a reference.

The selection of NC DIS events is based on the identification
of the scattered positron as the most energetic compact calorimetric deposit in the
SpaCal  with an
energy $E_e'>7.5~\rm GeV$ and a polar angle $156\deg<\theta_e'<175\deg$.
The energy weighted radius of this cluster 
is required to be less than $4$~cm, as expected for 
an electromagnetic shower. The cluster must be geometrically
associated with a track candidate in the BDC.
%
The \mbox{$z$-coordinate} of the primary event vertex is required to be 
within $\pm 35~\rm cm$ 
of the nominal position of the interaction point.

The remaining clusters in the calorimeters and the charged tracks are 
combined to reconstruct the hadronic final state, using an algorithm
which 
avoids double counting of 
energy \cite{Peez:2003zd,Portheault:2005uu}. The total longitudinal energy balance, 
determined as 
the difference of the total energy $E$ and the longitudinal component of the total momentum $P_z$, 
calculated from all detected particles including the scattered positron, must satisfy 
$45 < E -P_{z} < 65~\gev$. This requirement reduces contributions of DIS events with 
hard initial state photon 
radiation. For the latter events, the undetected photons propagating in the negative $z$ direction 
lead to values of $E -P_{z}$ significantly lower than the expected value $2E_e=55.2\gev$. 
After this selection the contribution from photoproduction  is negligible
as estimated using Monte Carlo simulations.

The kinematic region covered by this analysis is defined by 
\begin{center}
$5<Q^2<100~\gevsq$~~~{\rm and}~~~$0.2<y<0.7$\,, 
\end{center} 
where $y=Q^2/(s \cdot x_{\rm Bj})$ 
quantifies the inelasticity of the interaction. These two variables are reconstructed from the four 
momenta of the scattered positron and the hadronic final state particles using the electron-sigma 
method \cite{Bassler:1995uq}.

Jet finding is performed in the Breit frame. The boost from the 
laboratory system is determined by $Q^2$, $y$ and by the azimuthal angle 
of the scattered positron. Particles 
of the hadronic final state are clustered into jets using the inclusive 
$k_T$ algorithm 
\cite{Ellis:1993tq,Catani:1993hr} 
with the massless $P_T$ recombination scheme and with the distance 
parameter in the $\eta-\phi$ plane  $R_0 = 1$. 
The cut $-1<\eta_{\rm Lab}^{\rm jet}<2.5$, where 
$\eta_{\rm Lab}^{\rm jet}$ is the jet pseudorapidity in the 
laboratory frame, ensures that jets are contained 
within the acceptance of the LAr calorimeter.
The transverse energy 
of jets in the Breit frame is required to be above $5$~GeV. 
Jets are ordered by decreasing transverse 
momentum $P_{T}$ in the Breit frame, which is identical to the 
transverse energy $E_T$ for massless 
jets. The jet with highest $P_T$ is referred to as the ``leading jet".

Three jet samples are defined: the inclusive jet sample contains all 
jets which satisfy the jet selection criteria; the 2-jet and 3-jet
samples contain events with at least 2 and 3 jets, respectively.
In addition, to avoid regions of phase space where fixed order perturbation 
theory is not reliable \cite{Frixione:1997ks}, 2-jet events are accepted 
only if the invariant mass $M_{12}$ of the two leading jets exceeds 18 GeV.
The same requirement is applied for 3-jet events such that the 3-jet sample 
is a subset of the 2-jet sample.

The selection criteria are summarised in Table \ref{tab:selection}.
 The final inclusive jet sample contains $164522$ events with $230140$ jets.
 The 2-jet and 3-jet samples contain $31550$  and $4879$ events
 respectively.
	

\subsection{Determination of the jet cross sections}
\label{sec:mc}

In order to extract the jet cross sections at hadron level,
the experimental data are corrected bin-by-bin for 
effects of limited detector acceptance and resolution and
for QED radiation effects.
The following leading order Monte Carlo event generators are used for the correction procedure: 
\mbox{DJANGOH~\cite{Charchula:1994kf}}, which uses the Colour Dipole Model with QCD matrix 
element corrections as implemented in ARIADNE~\cite{Lonnblad:1992tz}, and 
RAPGAP~\cite{Jung:1995gf}, based on 
QCD matrix elements matched with parton showers in leading log approximation. 
The effects of QED radiation are included using the HERACLES   
\cite{Kwiatkowski:1992es} program interfaced with RAPGAP and DJANGOH.
In both Monte Carlo generators the hadronisation is modelled with Lund string 
fragmentation~\cite{Andersson:1983ia}. 
The generated events are passed through a GEANT3~\cite{Brun:1987ma} based 
simulation of the H1 apparatus 
and are reconstructed using the same program chain as for the data. 
Both the RAPGAP and DJANGOH simulations provide a good overall description
of the shapes of all relevant data distributions.
To further improve the agreement between Monte Carlo and data, a reweighting 
as a function of $Q^2$ and $P_T$ of the leading jet is applied 
to the Monte Carlo events.

The bin dependent correction factors are determined 
from Monte Carlo simulations as the ratios of the cross sections 
obtained from particles at hadron level without QED radiation
to the cross section calculated using reconstructed particles
and including QED radiation effects.
The mean values of the correction factors determined by RAPGAP and DJANGOH
are used, and half of the difference is assigned as a model 
uncertainty.  The typical value of these factors is between 
$1.2$ and $1.4$.

The binnings in $Q^2$, $P_T$ and $\xi$ used to measure the 
jet cross sections are given in Table \ref{tab:bins}.
The bin purities, defined as the fraction of events reconstructed
in a particular bin that originate from that bin on hadron level, 
is found to be typically $70\%$ and larger than $50\%$ in all analysis bins. 
The bin stabilities, 
defined as the fraction of events originated 
from a particular bin on hadron level that are reconstructed in that bin,
is typically $60\%$ and
larger than $40\%$ in all analysis bins.

\subsection{Experimental uncertainties}
\label{sec:sys}

Several sources of experimental uncertainties are considered.
The systematic uncertainties of the jet cross sections
are determined by propagating the corresponding estimated 
measurement errors through the full analysis:

\begin{itemize} 
 \item
 
 The relative uncertainty of the positron energy calibration is better 
 than $1\%$. The absolute uncertainty of the positron polar angle 
 is about  $1$~mrad. 
 Uncertainties in the positron reconstruction affect the event kinematics and thus the 
 boost to the Breit frame. This in turn leads to a relative error of up to $2\%$ on the 
 jet cross section for each of the two sources.
 
\item
 The relative uncertainty on the energy of the reconstructed hadronic final 
state as well as of the jet energy is estimated to be $2\%$.
It is dominated by the
uncertainty of the hadronic energy scale of the LAr calorimeter. 
Two different calibration methods are used for 
jet transverse momentum below and above $10$~GeV, respectively.
The resulting uncertainty on 
the cross sections is typically in the range of $4\%$ to $10\%$. 
The uncertainty of the SpaCal hadronic 
energy scale of $7\%$ contributes less than 
$1\%$ to the uncertainty of the cross section.

\item 
  The model dependence of the detector correction factors is estimated as described in 
section~\ref{sec:mc}. 
It reflects the sensitivity of the detector simulation to the details of the 
model, especially the parton showering and its impact 
on the migration between adjacent bins in 
$P_T$. The model dependence is below $10\%$ in most of the bins and typically $4\%$.

\item 
       The luminosity measurement uncertainty leads to an 
       overall normalisation error of the jet cross
       sections of $1.5\%$.
\end{itemize}

The uncertainty of the luminosity 
measurement is assumed to be fully correlated between the bins. 
The remaining sources of systematics, namely the positron energy scale 
and polar angle, the hadronic final state energy scale and the model dependence are 
assumed to be equally shared between correlated and uncorrelated parts.

The dominant experimental uncertainties on the jet cross sections arise 
from the model dependence of 
the data correction and from the LAr hadronic energy scale uncertainty. 
The individual contributions 
are added in quadrature to obtain the total systematic uncertainty.

\section{NLO QCD Calculations}
\label{sec:nlo}

Reliable quantitative predictions of jet cross sections in DIS require the perturbative 
calculations to be performed at least to NLO in the strong coupling. 
The NLO calculations are used for comparison to data and for the $\alps$ extraction.
By using the 
inclusive $k_T$ jet algorithm, the observables in the present analysis are 
infrared and collinear safe.
Application of this algorithm in the Breit frame allows the
initial state singularities to be absorbed in the definition
of the proton parton densities, as needed for the calculation
of factorised jet cross sections~\cite{Webber:1993bm}.

 Jet cross sections are predicted at the parton level 
using the NLOJET++ program~\cite{Nagy:2001xb} at NLO 
in the strong coupling using the same jet definition as in the data analysis.
When comparing data and theory predictions 
the  strong coupling 
is taken to be $\alpha_s(M_Z) = 0.118$ 
at the $Z^0$ boson mass and 
is evolved as a function of the renormalisation scale 
with two loop precision. 
No QED radiation is included in the calculations, 
but the running of the electromagnetic coupling with $Q^2$ 
is taken into account in the theoretical predictions.
The calculations are performed in the $\overline{\mbox{\rm MS}}$ scheme \cite{Bardeen:1978yd}
for five massless quark flavours. The parton 
density functions (PDFs) of the proton are taken 
from the CTEQ6.5M set~\cite{Tung:2006tb}. The 
factorisation scale $\mu_f$ and 
the renormalisation scale $\mu_r$ are taken to be 
$\sqrt{(Q^2 + P_{T\raisebox{0pt}[0pt][0pt]{,\,}\text{obs}}^{2})/2}$ for the NLO predictions, 
with $P_{T,\rm obs}$ denoting the $P_T$ of the 
jet for inclusive jet cross sections
and the average transverse momentum of the two leading jets $\Ptav$
for the  2-jet and 3-jet cross sections.
This choice of the scales is motivated by the presence of two hard scales
in jet production in DIS, $P_T$ and $Q$, the latter being smaller
in most of the analysis bins.
The calculations were also performed using  $\mu_r=P_{T,\rm obs}$.
With this choice of renormalisation scale the NLO QCD prediction
decreases by $10-20\%$ at lowest $Q^2$ and $P_T$ and
is disfavoured by the data.

Hadronisation corrections are calculated for each bin 
using Monte Carlo event generators 
DJANGOH and RAPGAP which implement different models 
for parton showering. 
These  corrections are determined as the ratio of the 
cross section at hadron level to the cross 
section at the parton level after parton showers. 
It was verified that the parton level jet cross sections
obtained from DJANGOH and RAPGAP are in
agreement with those from the NLO calculation
within the systematic uncertainties considered here.
The hadronisation correction factors are determined
as the average of values obtained from DJANGOH and RAPGAP.
Half of the difference is assigned as hadronisation uncertainty
and included as a part of the theoretical uncertainty.
For inclusive and 2-jet cross sections the hadronisation 
correction factors differ typically by less than $10\%$ 
from unity and agree at the level of $2$ to $5\%$ between the 
two Monte Carlo simulations. 
For 3-jet cross sections, as well as for the ratios 
of 3-jet to 2-jet cross sections, the hadronisation 
correction factors differ from unity by about 
$20\%$ with up to  $10\%$ difference between the two  MC models. 

The dominant theoretical uncertainty is 
related to the missing higher orders in the perturbative calculation,
and is conventionally estimated by separately varying the scales 
$\mu_f$ and $\mu_r$ by factors in the arbitrary range $0.5$ to $2$. 
The contributions from the two scale variations are similar and are
added in quadrature to obtain the total scale dependence uncertainty. 
The uncertainty originating from the 
PDFs is taken into account for the $\alpha_s$ extraction
using the variations of the CTEQ6.5M set of parton densities.

\section{Cross Section Measurements}
\label{sec:res}

In the following, the differential cross sections, 
corresponding to the phase space given in 
Table~\ref{tab:selection}, are presented for inclusive jet, 
2-jet and 3-jet production at hadron level.
Ratios of 3-jet to 2-jet hadron level cross sections
are also presented.
The measurements are shown in 
Tables \ref{tab::jet_Q2} to \ref{tab::Ratio_Q2Pt} 
and Figs.~\ref{incljets} to \ref{ratios322}.

\subsection{Inclusive jet cross section}

The measured inclusive jet cross section, corrected for detector and
radiative QED effects, is presented as a function of $Q^2$ and $P_T$
of the jet, as single 
differential distributions in Figs.~\ref{incljets}(a,b)
 and double differentially in Fig.~\ref{incljets-double}.
Each jet which satisfies the jet selection criteria described in 
section~\ref{sec:jetsel} enters these distributions.

The measurements are well described by NLO QCD predictions  corrected for hadronisation 
effects as explained in section~\ref{sec:nlo}. 
%
%
The theoretical uncertainty, dominated by the scale variation,
reaches $30\%$ for the lowest $Q^2$ and $P_T$ bins and decreases 
to $10\%$ for the highest $Q^2$ and $P_T$ values. 
The relative contribution of hadronisation 
corrections to this error is small. The PDF uncertainty 
is about $6\%$ at the lowest $Q^2$ and $P_T$ 
and decreases to $2\%$ for the highest $Q^2$.

\subsection{2-jet and 3-jet cross sections}
\label{sec:23jets}

\begin{figure}[ht]
\centering
\epsfig{file=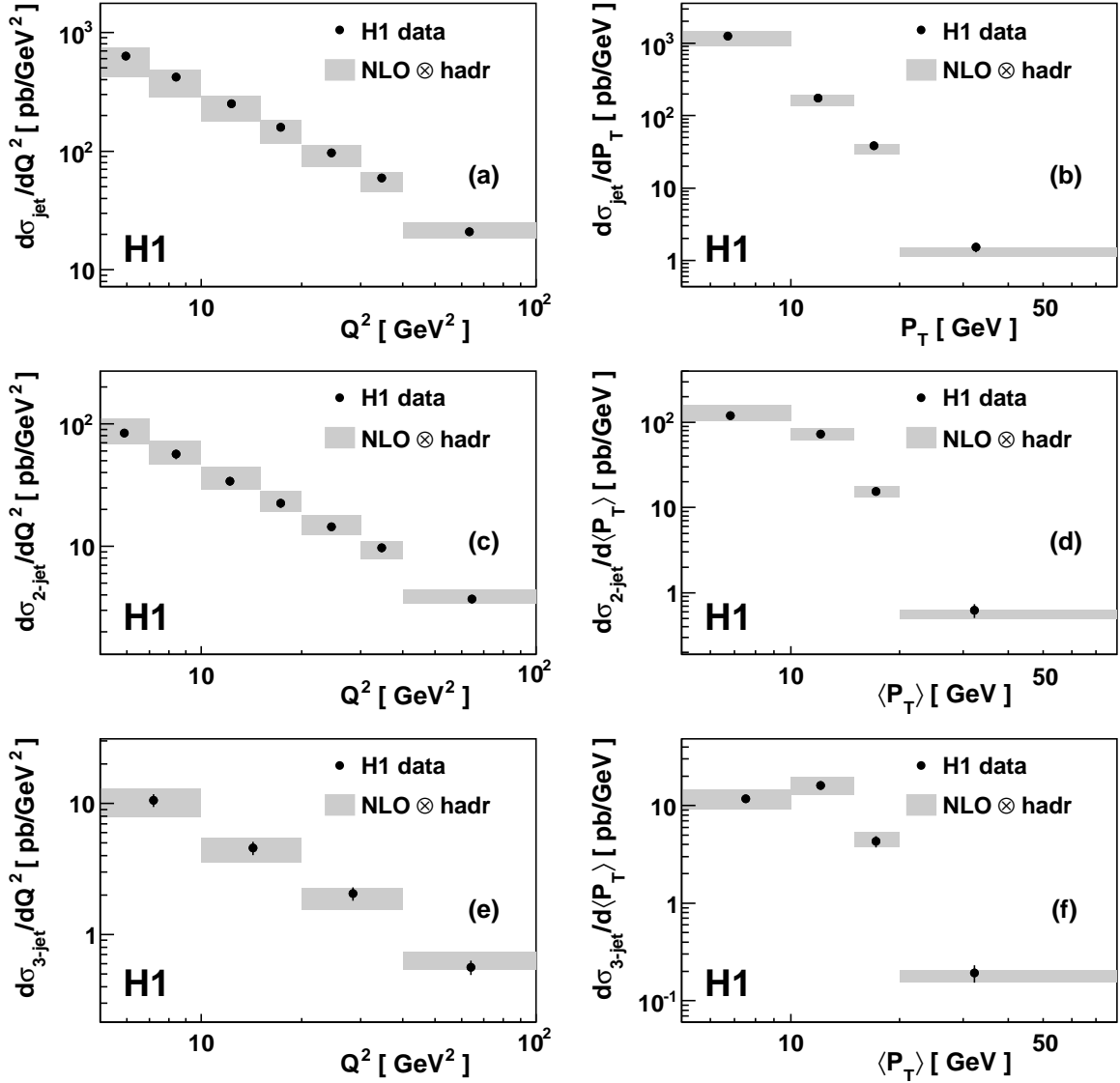,width=160mm}
\caption{Inclusive jet cross sections 
 $d\sigma_{jet}/dQ^2$ (a) and $d\sigma_{jet}/dP_T$ (b), 
 2-jet cross sections 
 $d\sigma_{2-jet}/dQ^2$ (c) and  $d\sigma_{2-jet}/d\Ptav$ (d)
 and 3-jet cross sections 
 $d\sigma_{3-jet}/dQ^2$ (e) and  $d\sigma_{3-jet}/d\Ptav$ (f), 
 compared with NLO QCD predictions corrected for hadronisation. 
 The error bars show the total experimental uncertainty, 
 defined as the quadratic sum of the statistical and systematic uncertainties.
 The points are shown at the average values of $Q^2$, $P_T$ or $\Ptav$
 within each bin.
 The NLO QCD predictions  are shown together with the
 theory uncertainties associated with the scale uncertainties and
 the hadronisation (grey band).
}
\label{incljets}
\end{figure}

\begin{figure}[ht]
\centering
\epsfig{file=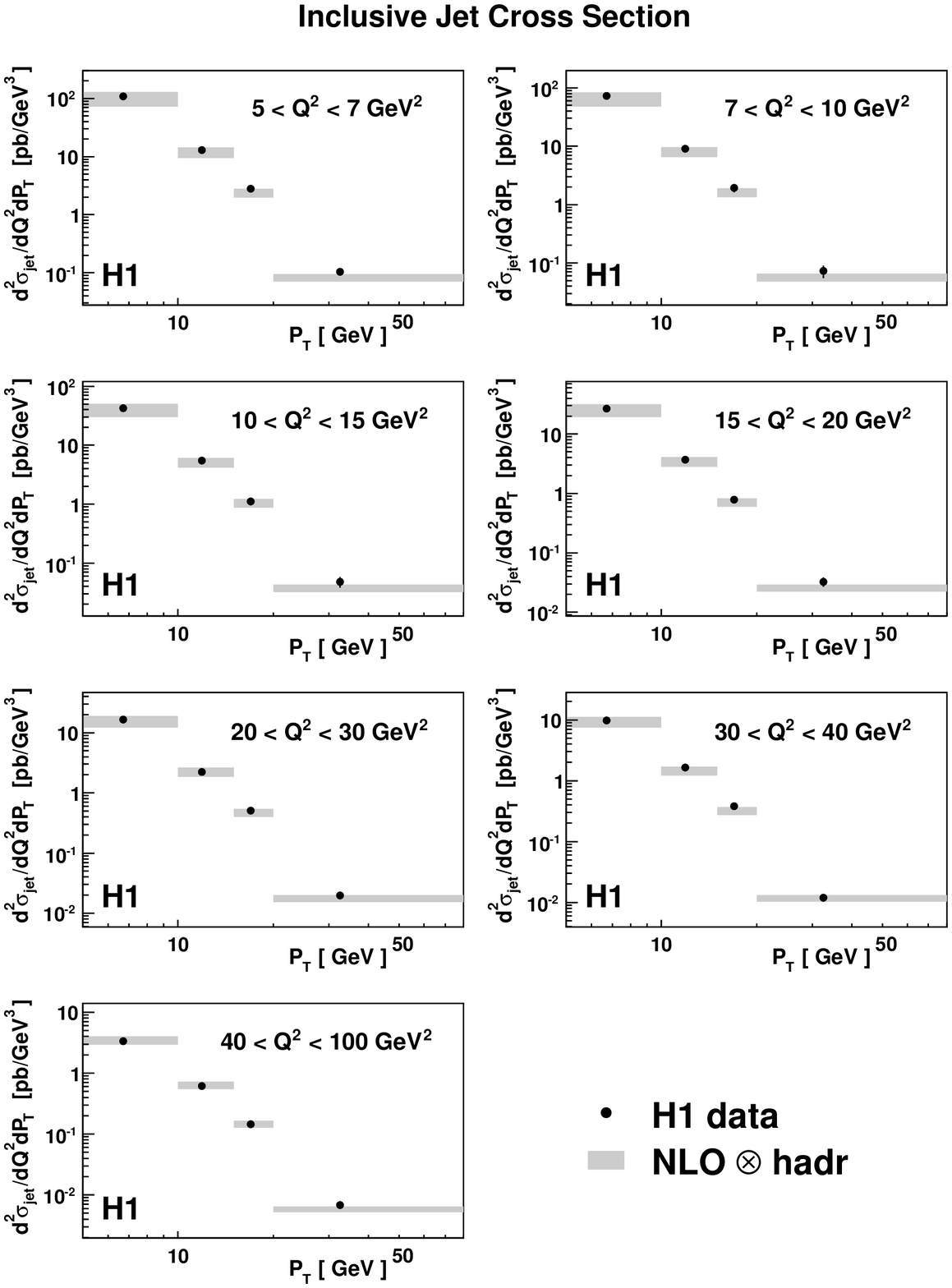,width=160mm}
\caption{Double differential inclusive jet cross sections as a function of
$Q^2$ and $P_T$, compared with NLO QCD
 predictions corrected for hadronisation.
 Other details are given in the caption to Fig.~\ref{incljets}.
 } 
\label{incljets-double}
\end{figure}

The measured single differential cross sections 
for 2-jet and 3-jet production as 
functions of $Q^2$ and of the average transverse momentum 
of the two leading jets 
$\left\langle P_T \right\rangle$ are shown in 
Figs.~\ref{incljets}(c,e) and Figs.~\ref{incljets}(d,f)
and are well described by the NLO QCD 
calculations corrected for hadronisation.
The relative uncertainties on the NLO QCD calculations of 2-jet, 3-jet and
inclusive cross sections are all of similar size.

\begin{figure}[ht]
\centering
\epsfig{file=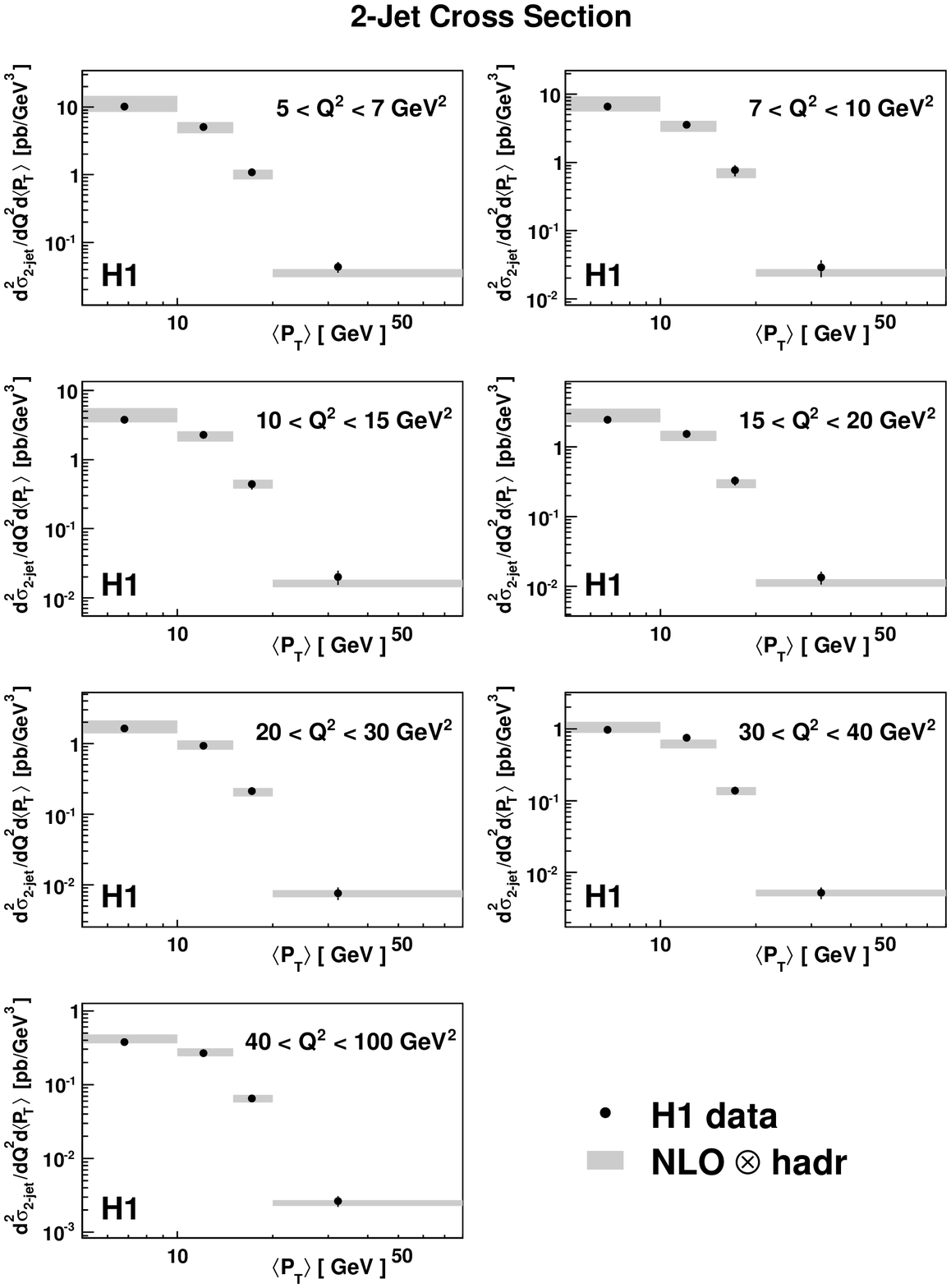,width=160mm}
\caption{Double differential 2-jet cross sections 
 as a function of $Q^2$ and $\Ptav$, compared with NLO QCD
 predictions corrected for hadronisation.
 Other details are given in the caption to Fig.~\ref{incljets}.
 } 
\label{dsdQ2dEtprime}
\end{figure}

\begin{figure}[ht]
\centering
\epsfig{file=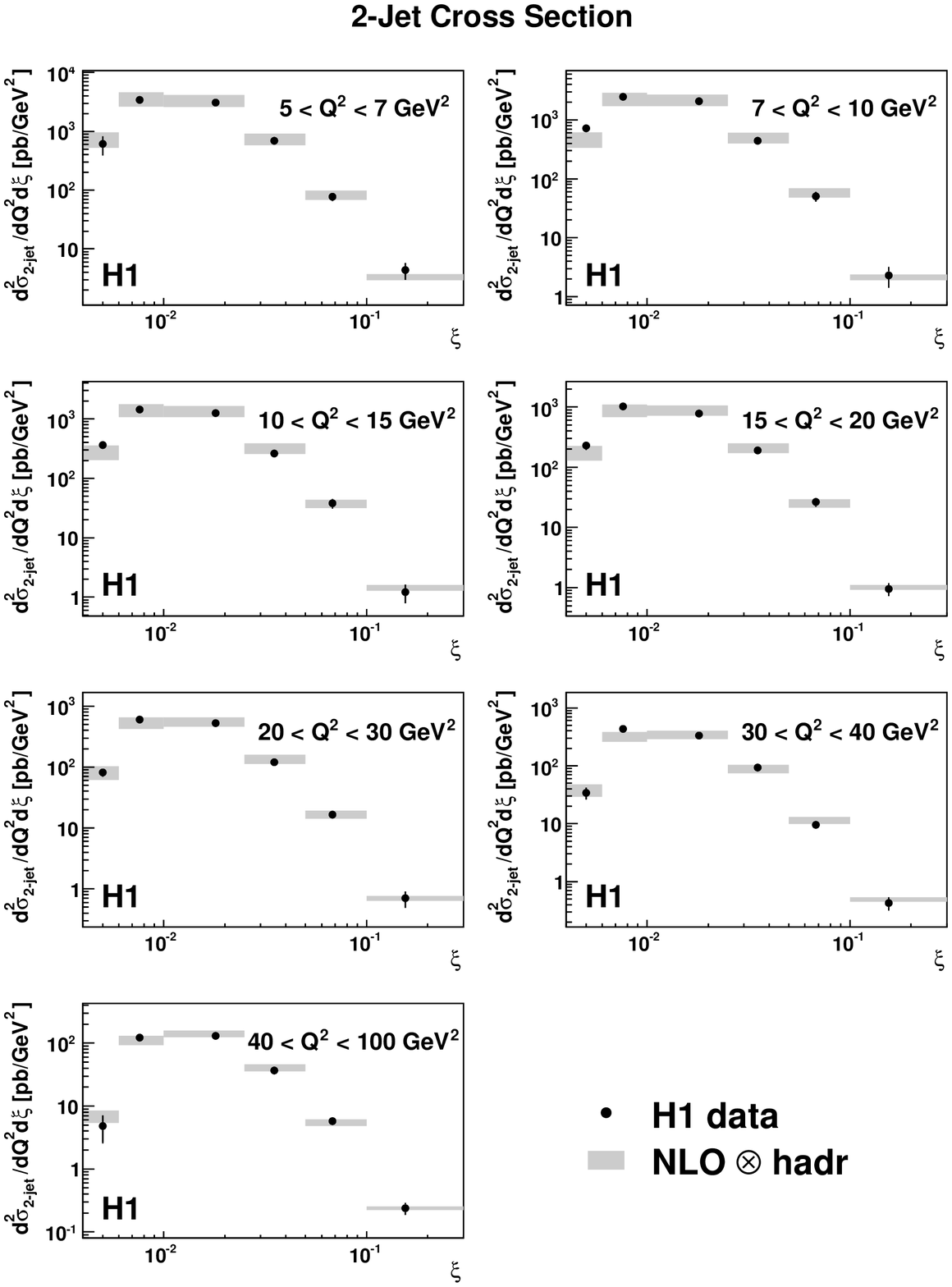,width=160mm}
\caption{Double differential 2-jet cross section, 
as a function of $Q^2$ and $\xi$, compared with NLO QCD
predictions corrected for hadronisation.
 Other details are given in the caption to Fig.~\ref{incljets}.
}
\label{dsdQ2dksi}
\end{figure}

\begin{figure}[ht]
\centering
\epsfig{file=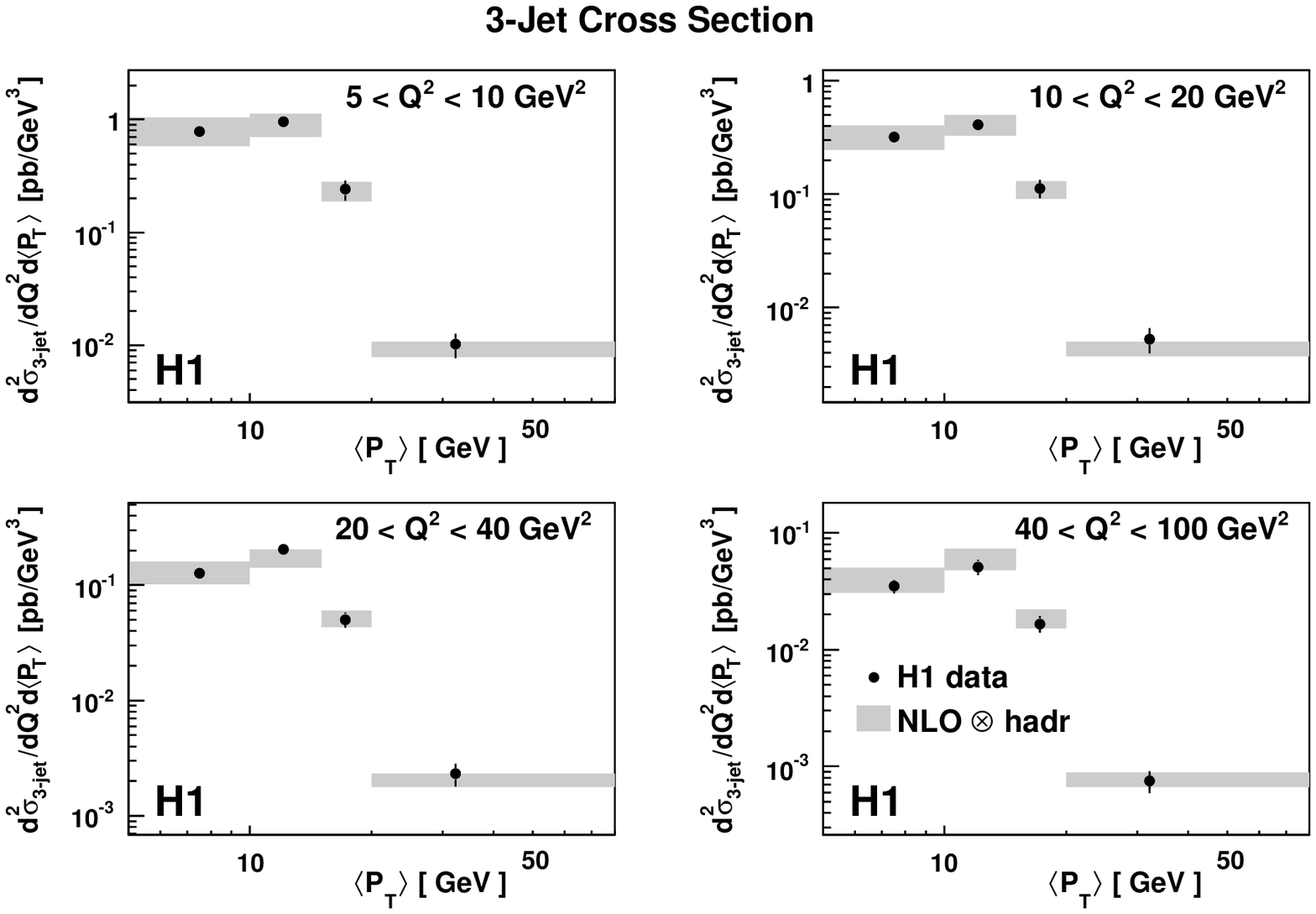,width=160mm}
\caption{Double differential 3-jet cross sections
 as a function of $Q^2$ and $\Ptav$, compared with NLO QCD
 predictions corrected for hadronisation.
 Other details are given in the caption to Fig.~\ref{incljets}.
 } 
\label{dsdQ2dEtprime3jets}
\end{figure}

The double differential 2-jet cross sections are presented 
in seven $Q^2$ bins 
as functions of the variables 
$\left\langle P_T \right\rangle$ and $\xi$ 
in Fig.~\ref{dsdQ2dEtprime} and 
Fig.~\ref{dsdQ2dksi} respectively. The 3-jet cross section 
is shown in four $Q^2$ bins 
as a function of $\Ptav$
in Fig.~\ref{dsdQ2dEtprime3jets}. 
The NLO QCD calculation 
provides an overall good description of the measured distributions 
within the quoted theoretical and experimental uncertainties.
Requirements in $P_T$ and $M_{12}$ suppress the cross section
at low $\xi$ where a rise is expected due to the increase of the 
gluon density.

The present results for the inclusive and the 2-jet cross sections
were compared to the previous H1 results in  \cite{Adloff:2000tq} and
\cite{Adloff:2002ew}, respectively. 
Taking into account the difference between proton beam energies and 
 differences in the kinematic region studied, the results are
found to be consistent with each other.

\begin{figure}[ht]
\centering
\epsfig{file=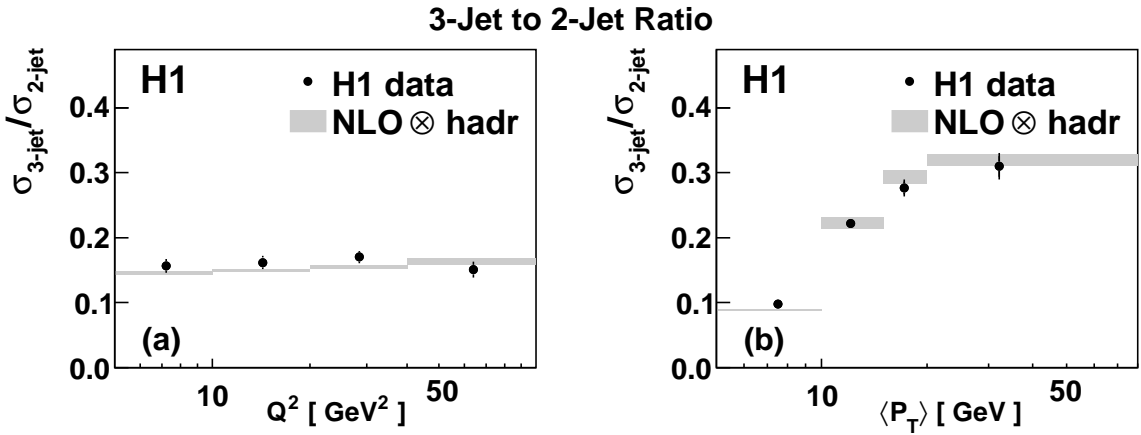,width=160mm}
\caption{Ratios of 3-jet to 2-jet cross sections
as a function of $Q^2$ (a) and $\Ptav$ integrated over the full $Q^2$ range (b) 
compared with NLO QCD predictions corrected for hadronisation.
 Other details are given in the caption to Fig.~\ref{incljets}.
}
\label{ratios32}
\end{figure}

\begin{figure}[ht]
\centering
\epsfig{file=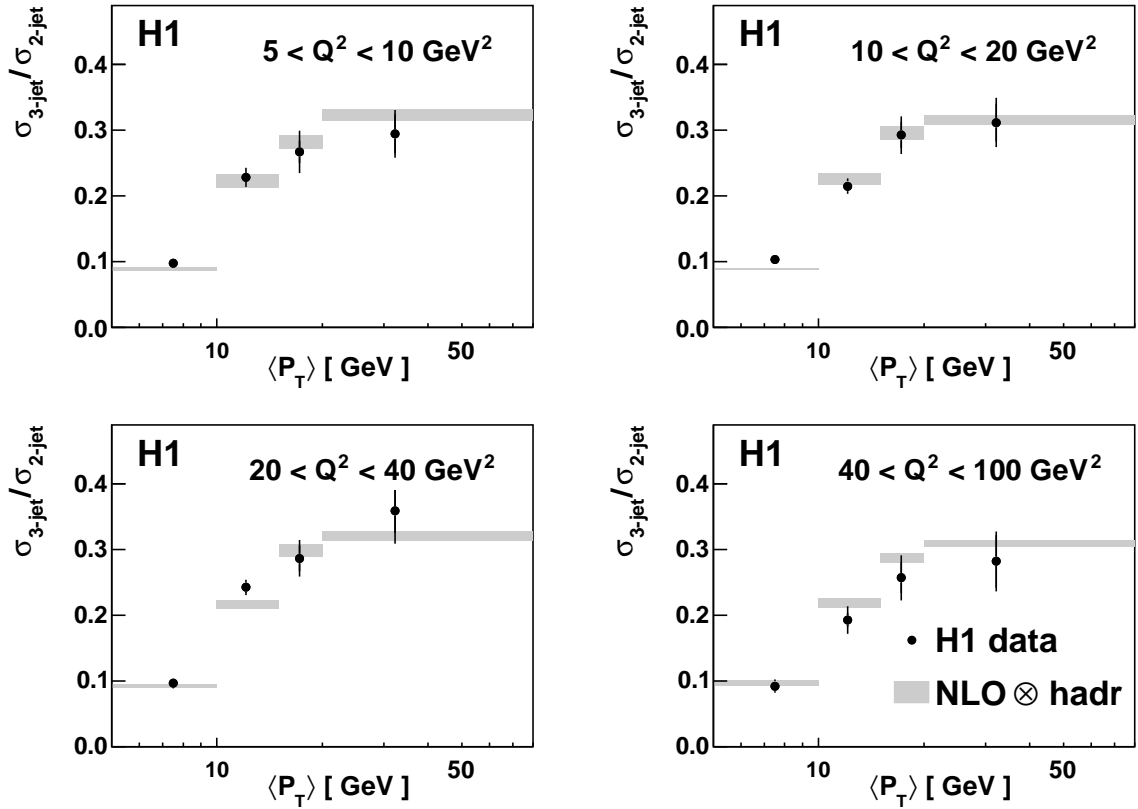,width=160mm}
\caption{Ratios of 3-jet to 2-jet cross sections
as a function of $\Ptav$ in four different $Q^2$ ranges
 compared with NLO QCD predictions corrected for hadronisation.
 Other details are given in the caption to Fig.~\ref{incljets}.
}
\label{ratios322}
\end{figure}

The 3-jet cross section normalised to the 2-jet cross section 
is presented in Fig.~\ref{ratios32} for single 
differential and in Fig.~\ref{ratios322} for double differential 
distributions.
 This observable benefits from cancellation of the normalisation 
uncertainties and reduction of the other systematic
uncertainties by about $50\%$. 
It is described by the NLO cross section except for the 
lowest $\Ptav$ bin, as seen in 
Fig.~\ref{ratios32}(b) and Fig.~\ref{ratios322}, and shows a 
reduced sensitivity to the renormalisation scale variation
which is done simultaneously for 2-jet and 3-jet cross sections. 
The ratio shows no significant dependence on $Q^2$ (Fig.~\ref{ratios32}(a)),
but increases with  $\Ptav$ (Figs.~\ref{ratios32}(b),\ref{ratios322})
mainly due to the increasing phase space.
%

\section{Extraction of the Strong Coupling}
\label{sect:Extract}

The QCD predictions for jet production depend on $\alps$ and on the 
PDFs of the proton. The strong coupling $\alpha_s$ is determined from 
the measured jet cross sections using the PDFs 
as obtained from inclusive DIS data and other measurements.

\subsection{Data and QCD predictions}
\label{sec:Data}

The $\alps$ determination is performed from individual bins 
of the double differential inclusive jet cross section,
$d^2\sigma/dQ^2dP_T$, and the 2-jet and 3-jet cross sections, 
$d^2\sigma/dQ^2d\Ptav$. 
Only bins are used in which the size of the
$k$-factor, defined as the ratio of the cross sections 
calculated in NLO and LO (both obtained with NLOJET++),
is below $2.5$.
The other bins most likely are affected by slow convergence of
perturbation series and exhibit a high scale dependence, up to $30\%$ at NLO.
The requirement that the $k$-factor be less
than $2.5$ corresponds to removing all points with 
$P_{T,\rm obs} < 10~$GeV for $Q^2<20~\GeV^2$ 
from the inclusive jet and the 2-jet cross sections 
and points with $10 < P_{\rm T}<15~$GeV 
for $Q^2<10~\GeV^2$ from the inclusive jet cross section 
\footnote{The bins removed from the $\alpha_s$ analysis
also correspond to the energy regime which is 
close to the $b$-quark mass threshold, where the five flavour 
massless approximation used in NLOJET++ is not expected to be valid.}.
In total, $62$ cross sections measurements are used for
$\alpha_s$ extraction: $22$ inclusive jet, $24$ 2-jet and 
all  $16$ 3-jet points.

QCD predictions of the jet cross sections are calculated 
as a function of $\alpha_s(\mu_r)$ with the FastNLO 
package \cite{Kluge:2006xs} using the CTEQ6.5M proton 
PDFs and applying the hadronisation corrections as 
described in section~\ref{sec:nlo}.

\subsection{The $\chi^2$ definition}
\label{sec:Fit}

Measurements and theory predictions are used to calculate 
a $\chi^2(\alps)$ with the 
Hessian \mbox{method \cite{Botje:1999dj}}, where parameters 
representing the systematic shifts of 
detector related observables, described in section \ref{sec:sys},
are left free in the fit. The shifts 
found by the  fit are consistent with the \textit{a priori} 
estimated experimental uncertainties. Due to 
different calibration strategies for jets with $P_T$ above and below $10$~GeV,
two different parameters are used for the hadronic final 
state energy scale for bins with 
$P_{T,\rm obs}<10~$GeV and $P_{T,\rm obs}\ge 10~$GeV.
The Hessian method used here takes into account correlations of 
experimental uncertainties and has also 
been used in global data analyses~\cite{Botje:1999dj,Barone:1999yv} 
and in previous H1 
publications~\cite{Aktas:2007pb, Adloff:2000dp}. 
The statistical correlations among 
the different bins and different observables are treated 
as described in~\cite{Aaron:2009vs,Gouzevitch:2008zzb}.
The experimental uncertainty of $\alpha_s$ 
is defined by the change in $\alps$ which gives an 
increase in $\chi^2$ of one unit with respect to the minimal value.

\subsection{Theory and PDF uncertainties}
\label{sec:Unc}

The theoretical uncertainty is estimated by the offset 
method as the difference between the value of $\alps$ from 
the nominal fit to the value when the fit is 
repeated with independent variations of different sources 
of theoretical uncertainties as described in section \ref{sec:nlo}. 
The resulting uncertainties due to the different 
sources are summed in quadrature. The up (or down) 
variations are applied simultaneously 
to all bins in the fit. The impact of hadronisation 
corrections and the factorisation 
scale uncertainty on $\alps$ typically amounts to $1\%$ to $2\%$ 
for each source. The largest uncertainty, of typically $8\%$, 
corresponds to the accuracy of the NLO approximation to the 
jet cross section estimated by varying the renormalisation scale as 
described in section \ref{sec:nlo}.

The uncertainty due to PDFs is estimated by 
propagating the CTEQ6.5M errors. The typical size of the resulting 
error is $2\%$ for $\alpha_s$ determined from the inclusive 
jet or 2-jet cross sections and $1\%$ when measured 
with the 3-jet cross sections. This uncertainty is twice as large 
as that estimated with the uncertainties given for the 
MSTW2008nlo90cl set \cite{Martin:2009iq} which in turn exceeds the 
difference between $\alps$ values extracted with the central sets 
of CTEQ6.5M, CTEQ6.6M~\cite{Nadolsky:2008zw} and MSTW2008nlo.

The CTEQ6.5M PDF parameterisation was obtained assuming $\alpha_s(M_Z)=0.118$.
In order to test whether this value of $\alpha_s(M_Z)$ biases the results 
obtained using the nominal method presented above, a method, similar to 
the one used in~\cite{Aktas:2007pb}, is employed using
the PDFs from the CTEQ6.6 series, which were
obtained assuming different values for $\alpha_s(M_Z)$.
The cross section as a function of the strong coupling 
is interpolated with a polynomial and this
interpolation is used to determine the best fit 
of the strong coupling to the data.
The result obtained with this alternative fit method 
is found to be compatible, well inside one standard deviation 
of the experimental error, with the value determined by the nominal method. 
Hence there is no indication for a bias due 
to the value of the strong coupling assumed for the CTEQ6.5M PDFs.

\subsection{Fit results}
\label{sec:Res}

\begin{figure}[ht]
\epsfig{file=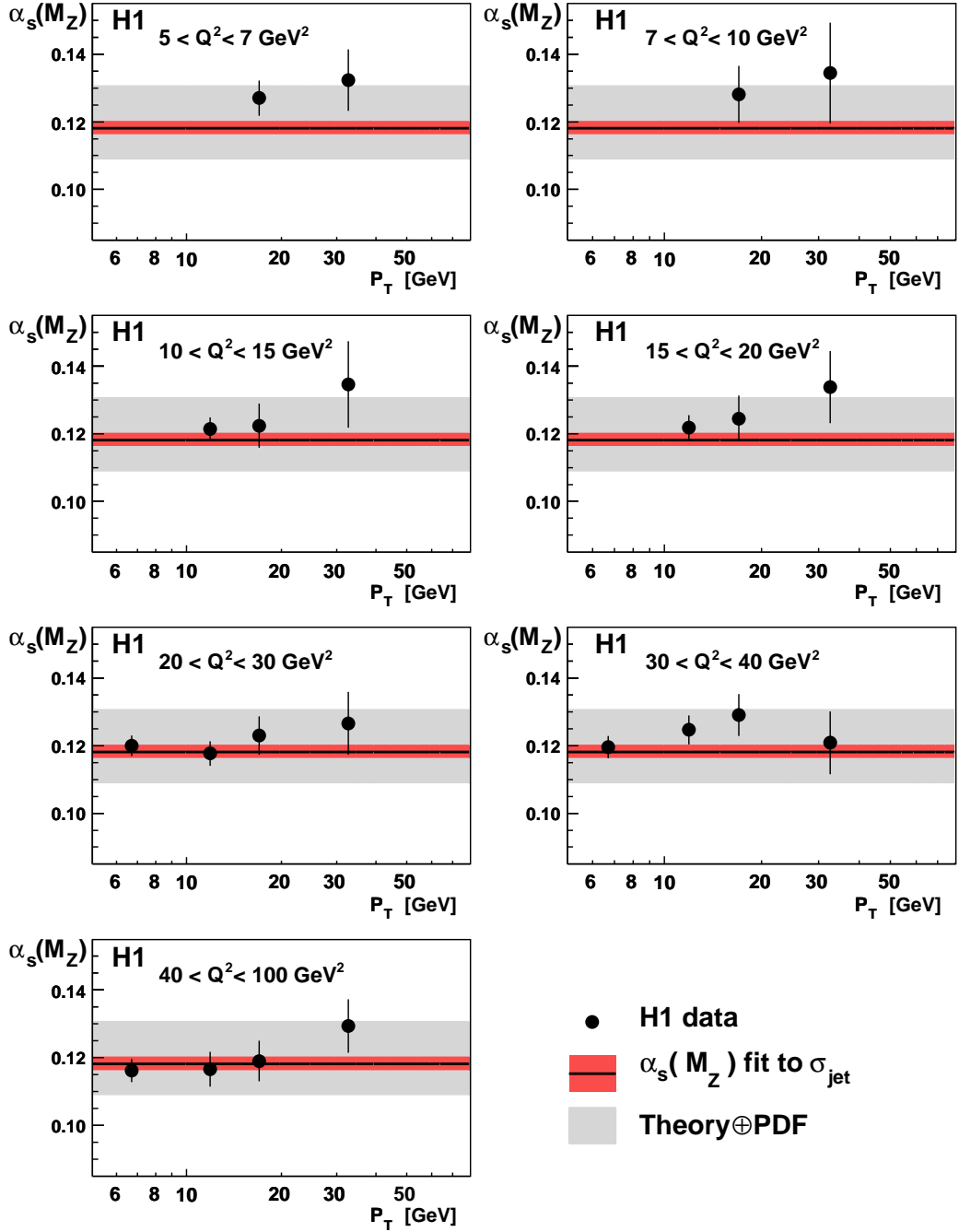,width=140mm}
\caption{Values of $\alpha_s(M_{Z})$ determined using 
the inclusive jet cross sections 
measured in $22$ bins in $Q^2$ and $P_T$ with the $k$-factor below $2.5$. 
The error bars denote the total experimental uncertainty of each data point. 
The solid line shows the two loop solution 
of the renormalisation group equation, $\alpha_s(M_Z)$, 
obtained from a simultaneous fit of all $22$ measurements 
of the inclusive jet cross sections. 
The inner band denotes the experimental uncertainty and 
the outer band denotes the squared sum of the PDF uncertainty  and the
theoretical uncertainties associated with the renormalisation
and factorisation scales  and the model dependence of the hadronisation 
corrections.
}
\label{fig:alphas1}
\end{figure}

\begin{figure}[ht]
\centering
\epsfig{file=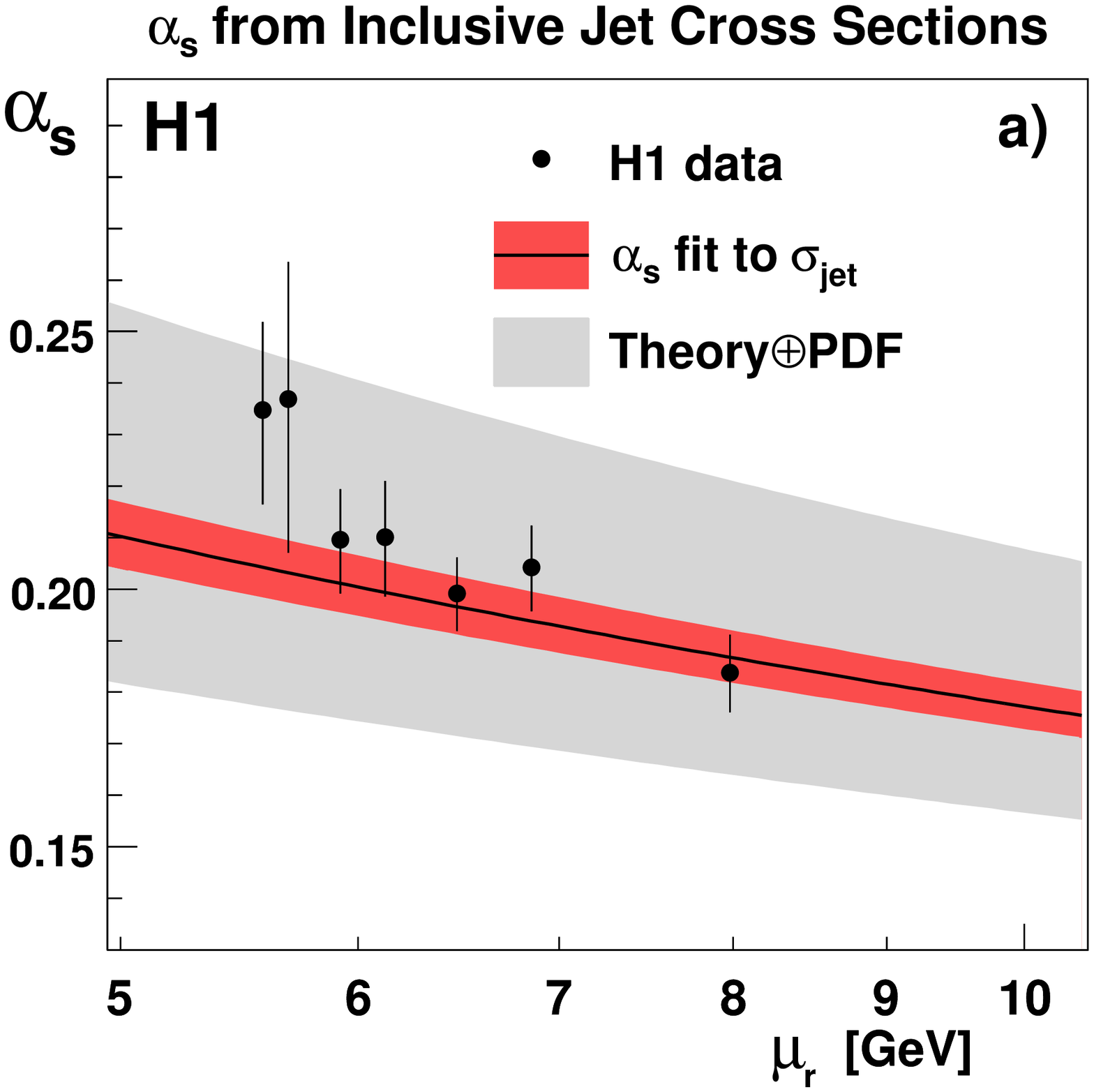,width=75mm}
\epsfig{file=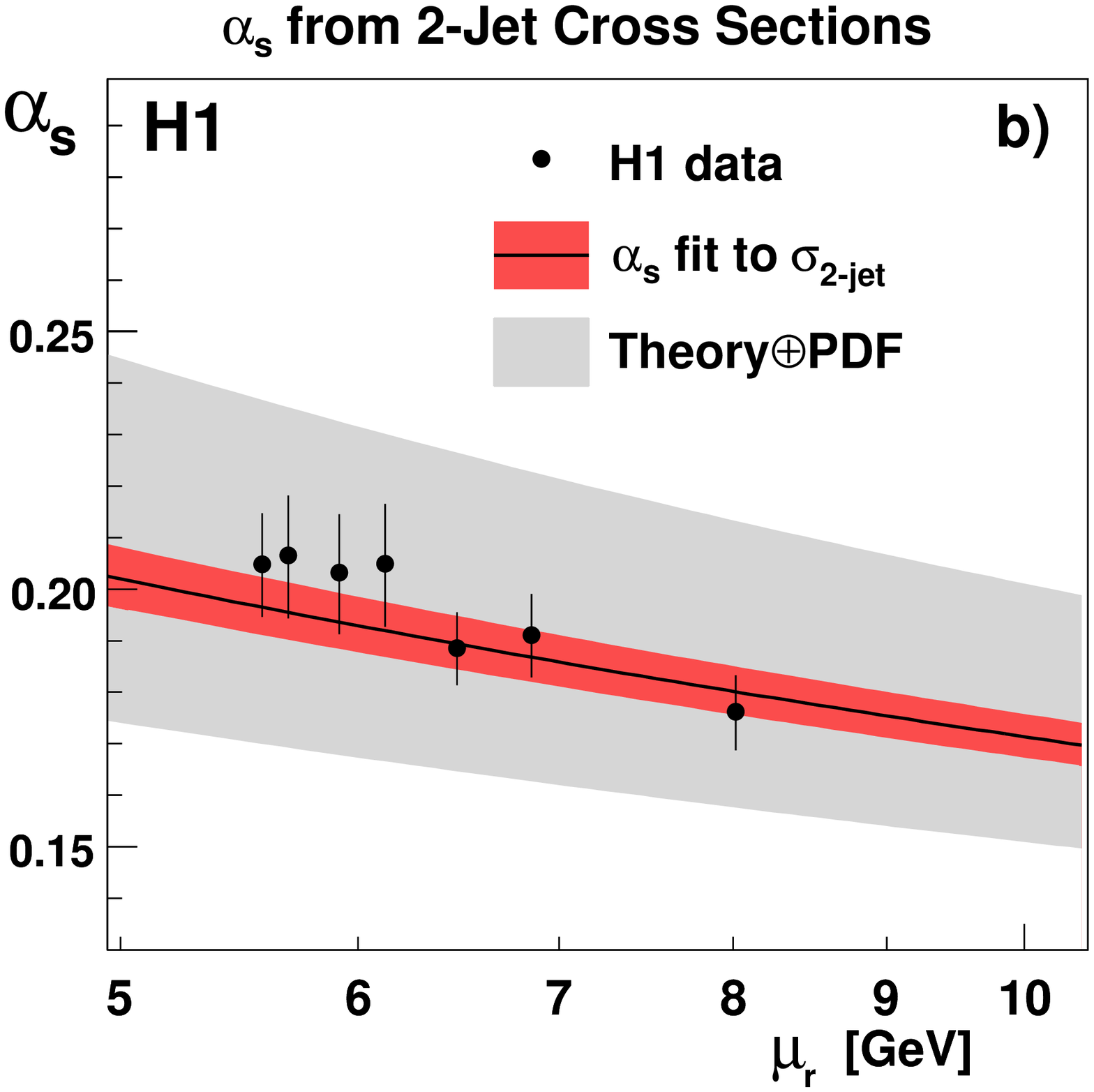,width=75mm}
\epsfig{file=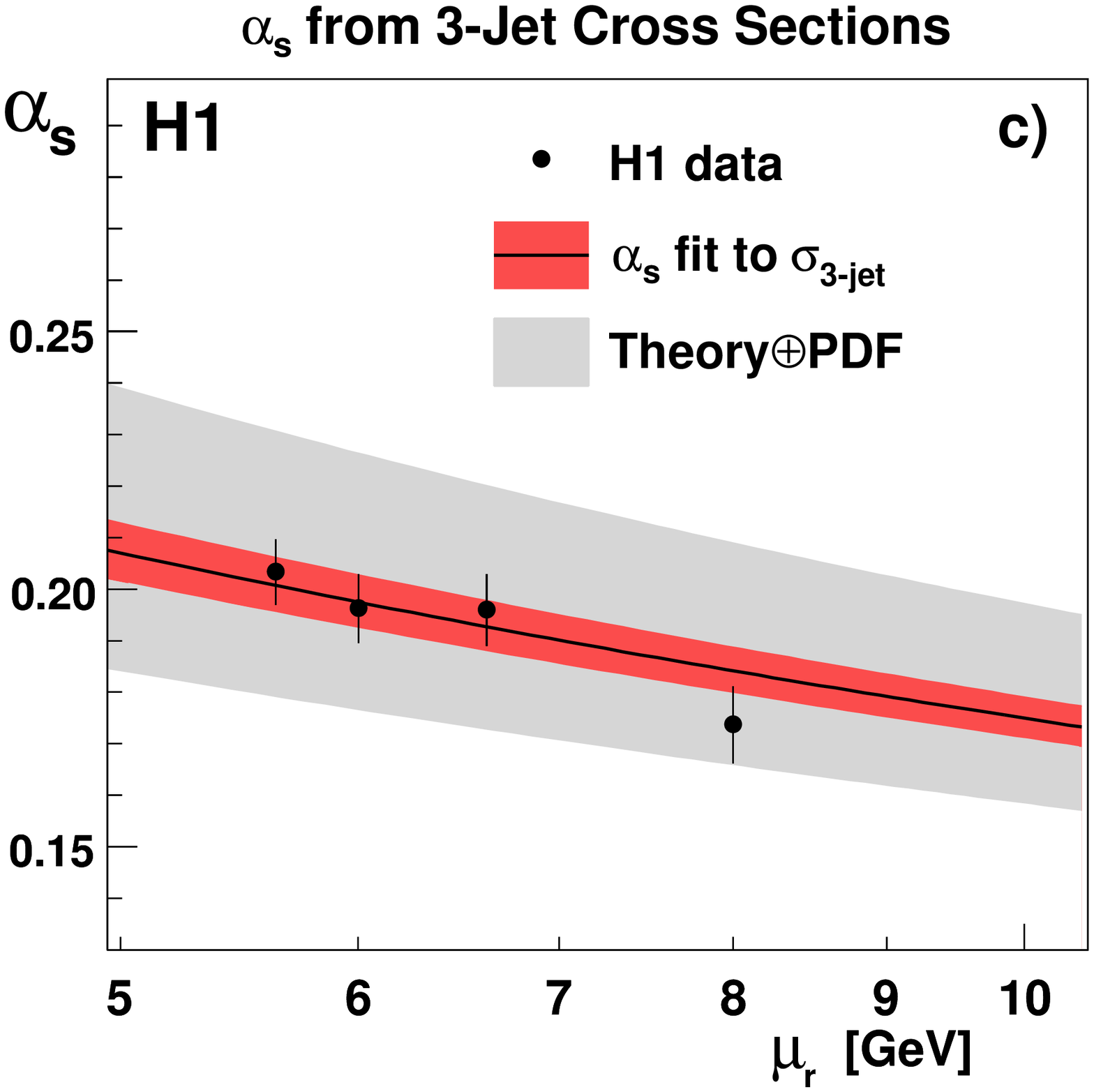,width=75mm}
\caption{Values of
 $\alpha_s(\mu_r=\sqrt{(Q^2 + P_{T\protect\raisebox{0pt}[0pt][0pt]{,\,}\text{obs}}^{2})/2})$
 extracted 
 from  inclusive jet cross section (a), 2-jet cross section (b)
 and 3-jet cross section (c).
 In each case, the solid lines show the two loop solution of the 
 renormalisation group equation obtained by evolving the corresponding 
 fitted value of $\alpha_s(M_Z)$, as summarized in table \ref{tab::Fits}, data rows 1-3. 
 Other details are given in the caption to Fig.~\ref{fig:alphas1}.
}
\label{fig:alphas2}
\end{figure}

\begin{figure}[ht]
\centering
\epsfig{file=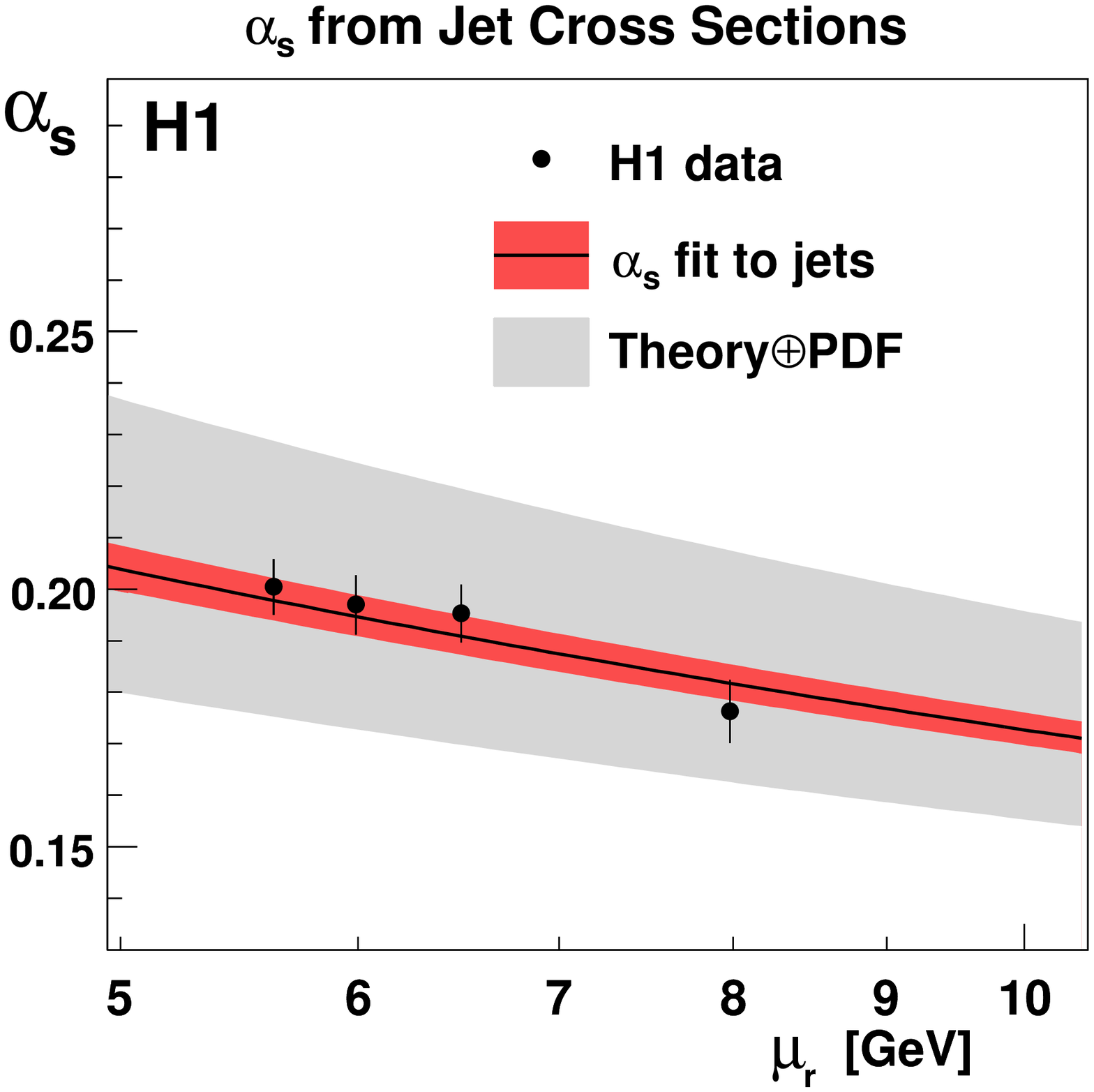,width=75mm}
\epsfig{file=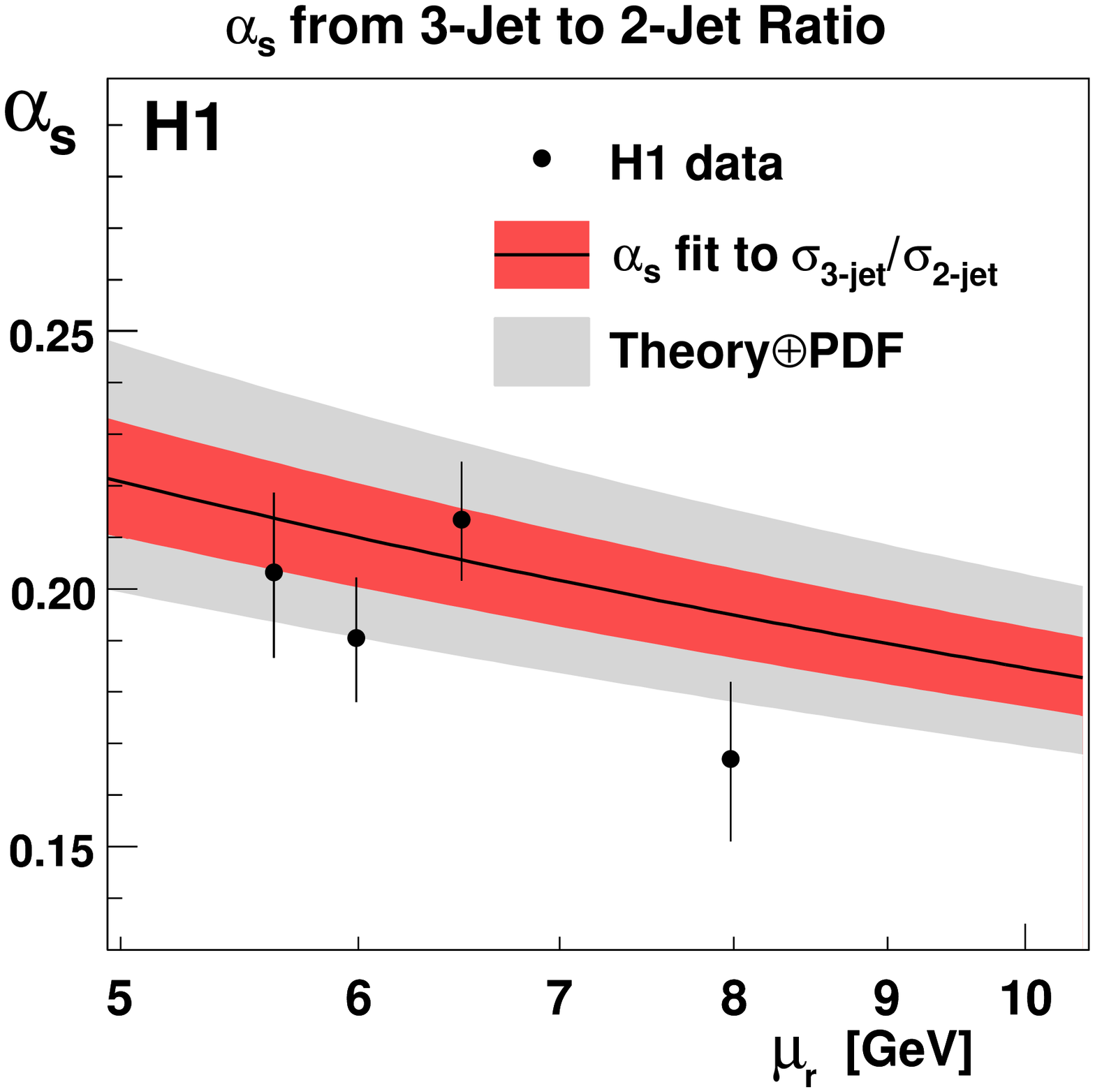,width=75mm}
\caption{Values of $\alpha_s(\mu_r=\sqrt{(Q^2 + P_{T\protect\raisebox{0pt}[0pt][0pt]{,\,} \text{obs}}^{2})/2})$ 
obtained by a simultaneous 
fit of all jet cross sections in each $Q^2$ bin (a) 
and of the ratio of 3-jet cross section to 2-jet cross section (b). 
The solid lines show the two loop solution of the 
renormalisation group equation obtained by evolving the $\alpha_s$ 
obtained from these measurements.
For (a) the value of $\alpha_s$ is extracted from a simultaneous fit 
of $62$ measurements of inclusive jet, 2-jet and 3-jet double differential
cross sections in bins of $Q^2$ and $P_T$ ($\Ptav$ for 2-jets and 3-jets)
with $k$-factor below $2.5$, see table \ref{tab::Fits}, 4th data row.
For  (b), $\alpha_s$ is extracted
from $14$ measurements of 
the 3-jet cross section normalised to 2-jet cross section, 
using only data points with  $k$-factor below $2.5$ 
for both the 3-jet and 2-jet cross sections, see table \ref{tab::Fits}, 5th data row. 
Other details are given in the caption to Fig.~\ref{fig:alphas1}.
}
 \vspace*{-68mm}

 \hspace*{-34mm}\bf (a) \hspace*{70mm} (b)
\vspace*{40mm}

\label{fig:alphas3}
\end{figure}

The fits of the strong coupling $\alpha_s $ are performed individually for each 
of the 62 cross section measurements 
as described in  section \ref{sec:Data}.
These measurements constrain the value of the 
strong coupling at the $Z^0$ mass, $\alpha_s(M_Z)$.
As an example,  $\alpha_s(M_Z)$ values determined from the inclusive jet
cross section are shown in Fig.~\ref{fig:alphas1}.
Different $\alpha_s(M_Z)$ values agree within experimental errors.
%
%
For each of the three observables and each $Q^2$ region, 
fits of $\alpha_s(M_Z)$ to all $P_T$ or $\langle P_T\rangle$ bins in that region are performed.
The fit results are evolved from $M_Z$ to the average scale $\mu_r$ in the
respective $Q^2$ regions, where the average renormalisation scale $\mu_r$ is calculated using NLO predictions. 
Figs.~\ref{fig:alphas2}(a-c) show the obtained $\alpha_s(\mu_r)$ values
for fits to inclusive, 2-jet and 3-jet cross-sections, respectively.
%
%
%
Also shown in these figures are QCD predictions $\alpha_s(\mu_r)$, derived
from common fits of $\alpha_s(M_Z)$ to all respective  $Q^2$ and $P_T$
or $\langle P_T\rangle$ bins. The results of these three common fits
are summarised in Table \ref{tab::Fits}. 
If the points with $k$-factor above $2.5$ are also included,
the $\alpha_s(M_Z)$ values obtained are changed by less than
one standard deviation of the total experimental uncertainty.
%
%
As the $\alpha_s(M_Z)$ measurements derived from inclusive, 2-jet or
3-jet observables agree within uncertainties in any of the $Q^2$
regions, they are combined within four $Q^2$ regions, taking into
account statistical and experimental systematic correlations. These
results, evolved from the scale $M_Z$ to the average $\mu_r$ in each
region, are shown in Fig.~\ref{fig:alphas3}(a). As compared to the
case of extracting $\alpha_s$ from only one variable, the
experimental uncertainties are reduced significantly.

Finally, all $62$ data points are used in a common fit of the strong 
coupling taking the correlations into account with 
a fit quality $\chi^2/{\rm ndf} = 49.8/61$:
\begin{equation}
\alpha_s(M_Z) = 0.1160 \pm 0.0014 (exp.) ^{+0.0093} _{-0.0077} (th.) \pm
0.0016 (\text{\scshape pdf}).
\label{eq:alps1}
\end{equation}
The experimental error on $\alpha_s(M_Z)$ measured with 
each observable typically amounts to $1.5\%$. 
The combination of different observables, even though 
partially correlated, gives rise to additional 
constraints on the strong coupling and 
leads to an improved experimental uncertainty of $1.2\%$. 
This error changes by at most 
$20\%$ when the scales, hadronisation factors and PDF 
parameterisation 
are changed within the limits defined by their uncertainties. 
The total error is dominated by the theoretical uncertainty 
of about $7\%$ mainly  due to scale variations. 
					
The determination of the strong coupling from the 
ratio of the 3-jet  
to the 2-jet cross section provides an alternative 
approach to combining the 
different cross section data. 
On the one hand the sensitivity of this observable to 
$\alpha_s$, which is $\mathcal{O}(\alpha_s)$, 
is reduced with respect to the 3-jet cross 
section, which is $\mathcal{O}(\alpha_s^2)$. 
On the other hand this observable 
benefits from reduced experimental and theoretical
uncertainties (see section \ref{sec:23jets}). The common fit of the 
strong coupling to the $14$ ratio points, for which the $k$-factors 
are below $2.5$ for both the 3-jet and 2-jet cross sections,
is given in Table \ref{tab::Fits}. 
The extracted $\alpha_s$  as a function of $\mu_r$ 
is shown in Fig.~\ref{fig:alphas3}(b). 
The experimental uncertainty on $\alpha_s$ increases  
 to $3\%$ with respect to the combined fit from the cross sections.
The theoretical uncertainty is reduced to $5\%$ and is dominated by the hadronisation 
uncertainty.

The strong coupling extracted from all $62$ data points (\ref{eq:alps1})
agrees well with that obtained from jet cross sections in the higher $Q^2$ range
between $150$ and $15000$ $\GeV^2$: 
%
$\alpha_s(M_Z) = 0.1168 \pm 0.0007 (exp.) ^{+0.0046} _{-0.0030} (th.) \pm
0.0016 (\text{\scshape pdf})~\text{\cite{Aaron:2009vs}}.
$
%
This agreement is remarkable, given the sensitivity of the NLO prediction to the 
renormalisation scale in the low $Q^2$ regime.
At high $Q^2$ \cite{Aaron:2009vs} the fit was done 
using the factorisation scale $Q$,
instead of $\sqrt{(Q^2 + P_{T\raisebox{0pt}[0pt][0pt]{,\,}\text{obs}}^{2})/2}$ used in this analysis.
However, as was shown in \cite{Aaron:2009vs}, the choice of factorisation
scale has only little impact on extracted value of $\alpha_s$
at high $Q^2$.
The value of $\alpha_s(M_Z)$ obtained in this analysis is also
consistent with the world 
averages $\alpha_s(M_Z)=0.1176 \pm 0.0020$ ~\cite{Amsler:2008zzb} and 
$\alpha_s(M_Z)=0.1184 \pm 0.0007$~\cite{Bethke:2009jm}.  

\begin{figure}[ht]
\centering
\epsfig{file=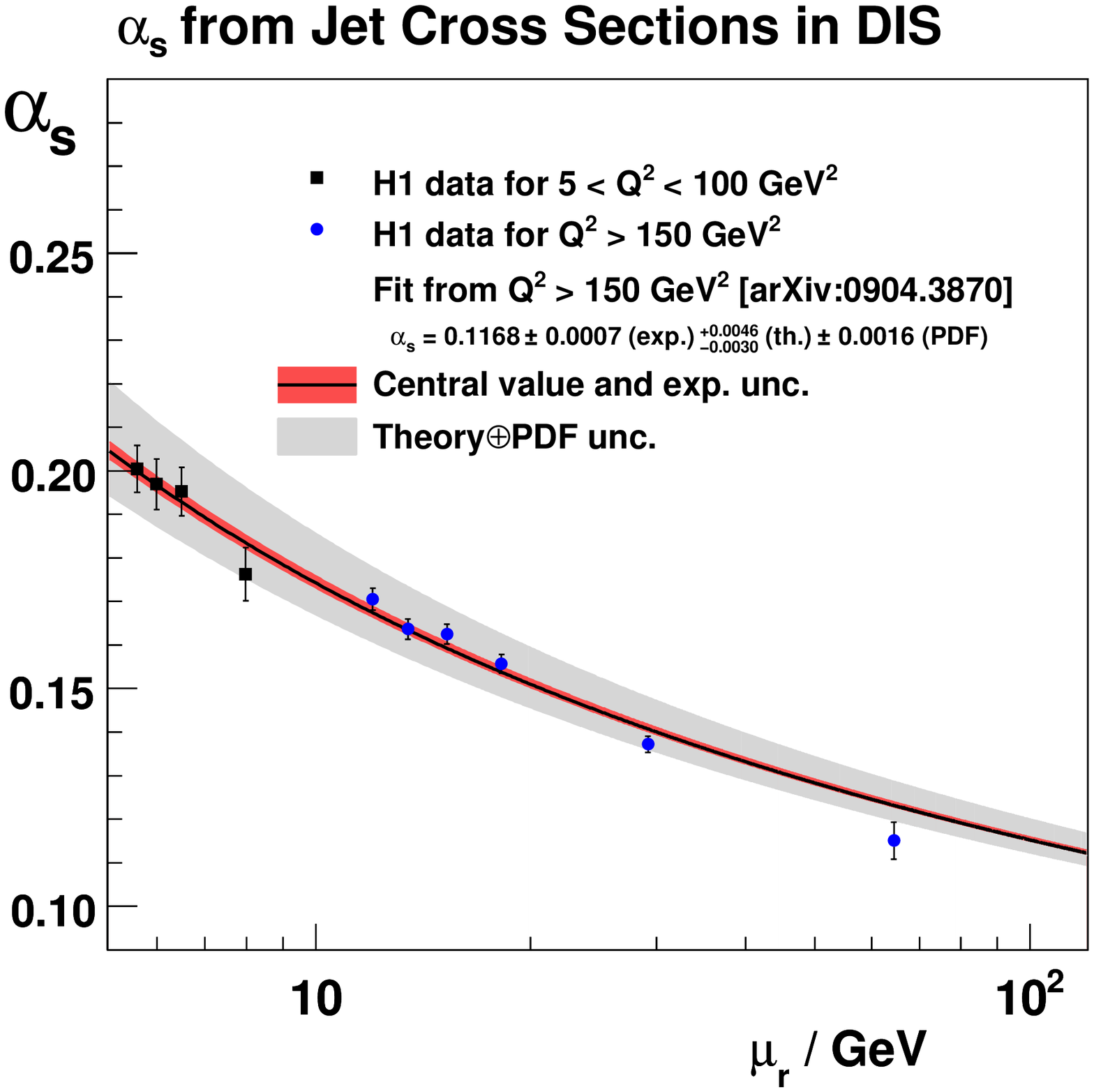,width=140mm}
\caption{Values of $\alpha_s(\mu_r=\sqrt{(Q^2 + P_{T\protect\raisebox{0pt}[0pt][0pt]{,\,} \text{obs}}^{2})/2})$ obtained by a simultaneous fit of all 
jet cross sections in each $Q^2$ bin of
this analysis (squares)  together with the fit in different 
bins at high $Q^2$ (circles) \cite{Aaron:2009vs}. 
The solid line shows the two loop solution of the 
renormalisation group equation obtained by evolving 
the $\alpha_s(M_Z)$ extracted from jets at high $Q^2$. 
Other details are given in the caption to Fig.~\ref{fig:alphas1}.
}
\label{fig:alphas4}
\end{figure}

The new low $Q^2$ measurement together with data from the 
high $Q^2$ analysis provides a   
test of the running of the strong coupling 
for $\mu_r$ between $6$ and $70$~GeV as 
illustrated in Fig.~\ref{fig:alphas4}.
A simultaneous fit of $\alps$ from low and high $Q^2$ data was also
performed. It did not lead to an improved precision with respect
to  the high $Q^2$ determination alone due to  the large
theoretical uncertainties at low $Q^2$.

	
\section{Conclusion}
\label{sect:Conclusion}

Measurements of the inclusive, 2-jet and 3-jet cross sections 
in deep-inelastic positron-proton scattering 
are presented in the range $5<Q^2<100~\gev^2$ and $0.2<y<0.7$. 
Jets are reconstructed using the inclusive $k_T$ algorithm
in the Breit frame and are required to have a minimum
transverse momentum of $5$~GeV.
Calculations at NLO QCD, corrected for hadronisation effects, provide a
good description of the single and double differential cross sections as
functions of the jet transverse momentum $P_T$, the boson virtuality
$Q^2$ as well as of the proton momentum fraction $\xi$. 
The precision of the measurements is typically $6$ to $10\%$.

The strong coupling $\alpha_s$ is determined from a 
fit of the NLO prediction to the measured jet cross sections. 
The dominant source of uncertainties is related to the 
renormalisation scale dependence, which is used to
estimate the effect of missing higher orders.
The extracted value of the strong coupling
$$
\alpha_s(M_Z) = 0.1160 \pm 0.0014 (exp.) ^{+0.0093} _{-0.0077} (th.) \pm
0.0016 (\text{\scshape pdf})
$$
is consistent with the value determined from high $Q^2$ jet cross sections.
Both measurements  test a running of 
the strong coupling for renormalisation scales $\mu_r$ between $6$ and $70$~GeV.



\section*{Acknowledgements}

We are grateful to the HERA machine group whose outstanding
efforts have made this experiment possible.
We thank the engineers and technicians for their work in constructing and
maintaining the H1 detector, our funding agencies for
financial support, the
DESY technical staff for continual assistance
and the DESY directorate for support and for the
hospitality which they extend to the non--DESY
members of the collaboration.
We thank Zoltan Nagy for fruitful discussions.

 \providecommand{\href}[2]{#2}\begingroup\raggedright
\endgroup

\newpage

\renewcommand{\le}{<}
\begin{table}[h]
	\centering
	\renewcommand{\arraystretch}{1.50}
  \begin{tabular}{| r | r | c | c |}
			\hline 
        NC DIS Selection &
	\multicolumn{3}{|c|}{$5 < Q^2 < 100~\GeV^2$ , $0.2 < y < 0.7$}  \\ \hline

     	Inclusive jet & $P_{T}>5~\GeV$        & &  \multirow{3}{*}{$-1.0 < 
        \eta_{\rm Lab}^{\rm jet} < 2.5$}  \\ 
        \cline{1-3}   2-jet         & 
        $P_{T}^{\rm jet1},~P_{T}^{\rm jet2}>5~\GeV$ & \multirow{2}{*}{$M_{12}>18~\GeV$}  & \\ 
        \cline{1-2}  3-jet         & 
        $P_{T}^{\rm jet1},~P_{T}^{\rm jet2},~P_{T}^{\rm jet3}>5~\GeV$ &   &  \\ \hline
  \end{tabular}
   \vspace{\baselineskip} 
	\caption{The NC DIS and jet selection criteria.}
  \label{tab:selection}
\end{table}

\begin{table}[h]
 \footnotesize
\centering
\begin{tabular}{ | c l | }
 \hline                 
                        &      \\
 bin number             &      \\            
 (inclusive and 2-jets) &    corresponding $Q^2$ range   \\[5pt]
 \hline
              &                                          \\[-8pt]
      1       &     $ \rm \  \ 5 \le Q^2< ~7~\GeV^2$     \\
      2       &     $ \rm \  \ 7 \le Q^2< 10~\GeV^2$  \\
      3       &     $ \rm    10 \le Q^2< 15~\GeV^2$     \\
      4       &     $ \rm    15 \le Q^2< 20~\GeV^2$     \\
      5       &     $ \rm    20 \le Q^2< 30~\GeV^2$     \\
      6       &     $ \rm    30 \le Q^2< 40~\GeV^2$     \\
      7       &     $ \rm    40 \le Q^2< 100~\GeV^2$  \\[3pt]    
 \hline                     &                               \\
 bin number                 &                               \\
 (3-jets and 3-jets/2-jets) &    corresponding $Q^2$ range  \\[5pt] 
 \hline
             &                                          \\[-8pt]
      I      &     $ \rm \ \ 5 \le Q^2 < 10~\GeV^2$  \\
      II     &     $ \rm  10 \le Q^2 < 20~\GeV^2$   \\
      III    &     $ \rm  20 \le Q^2 < 40~\GeV^2$   \\
      IV     &     $ \rm  40 \le Q^2 < 100~\GeV^2$   \\[3pt]
 \hline                     &                               \\
 bin letter   &    corresponding $P_T$ or $\Ptav$  range \\[5pt]
 \hline 
              &                                          \\[-8pt]
      a       &     $ \rm \ \ 5 \le P_T< 10~\GeV$   \\
      b       &     $ \rm  10 \le P_T< 15~\GeV$    \\
      c       &     $ \rm  15 \le P_T< 20~\GeV$    \\
      d       &     $ \rm  20 \le P_T< 80~\GeV$  \\[3pt]
 \hline                     &                               \\
 bin letter   &     corresponding $\xi$ range               \\[5pt]
\hline
              &                                          \\[-8pt]
      A       &     $ \rm  0.004 \le \xi< 0.006~$  \\
      B       &     $ \rm  0.006 \le \xi< 0.010~$  \\
      C       &     $ \rm  0.010 \le \xi< 0.025~$  \\
      D       &     $ \rm  0.025 \le \xi< 0.050~$  \\
      E       &     $ \rm \ \ 0.05 \le \xi< 0.1~$         \\
      F       &     $ \rm \ \ \ \ 0.1 \le \xi< 0.3$        \\[3pt]
 \hline
\end{tabular}
\caption{Nomenclature for the bins in 
negative four momentum transfer squared $Q^2$, 
jet transverse momentum $P_T$, 
average transverse momentum of the two leading jets $\Ptav$
and momentum fraction $\xi$ 
used in the following tables.}
\label{tab:bins}
\end{table}


\begin{table}[htp]
\centering
\tiny \sf
\renewcommand{\arraystretch}{1.50}
\begin{tabular}{|c || r || r | r | r | r || r | r | r | r || r | r|} 

\multicolumn{12}{c}{ } \\
\multicolumn{12}{c}{\normalsize Inclusive Jet Cross Section ${\rm \frac{d\sigma_{jet}}{dQ^2}}$ }\\
\multicolumn{12}{c}{ } \\
\hline
 & & &  & total & total &
\multicolumn{4}{|c||}{\underline{\hspace*{0.1cm}single contributions to 
                       correlated uncertainty\hspace*{0.1cm}} }
   &                                           hadronisation & hadronisation\\
bin   & cross        & statistical  & total        & uncorrelated & correlated & model  & electron     & electron   & hadronic    & correction    & correction    \\
      & section      & uncert.      & uncert.      & uncertainty  & uncert.    & uncert.& energy scale & polar angle&  energy scale &  factor    & uncertainty \\
    & $\rm [pb/GeV^{2}]$ & [\%]    & [\%]         & [\%]         & [\%]       &   [\%] & [\%]         & [\%]       & [\%] &       & [\%]\\
\hline
1 &$633\,   $&$0.7  $&$6.7  $&$4.7  $&$4.8  $&$1.7  $&$0.3  $&$0.4  $&$4.2  $&$0.88 $&$5.8$ \\ 
2 &$421\,   $&$0.6  $&$7.0  $&$4.9  $&$5.1  $&$2.2  $&$0.2  $&$0.4  $&$4.3  $&$0.89 $&$5.3$ \\ 
3 &$250\,   $&$0.6  $&$6.7  $&$4.7  $&$4.9  $&$1.4  $&$0.1  $&$0.7  $&$4.4  $&$0.89 $&$4.7$ \\ 
4 &$158\,   $&$0.7  $&$6.6  $&$4.6  $&$4.8  $&$1.4  $&$0.1  $&$0.7  $&$4.3  $&$0.90 $&$4.0$ \\ 
5 &$96.9\,    $&$0.7  $&$6.5  $&$4.5  $&$4.7  $&$1.3  $&$0.2  $&$0.5  $&$4.2  $&$0.91 $&$3.5$ \\ 
6 &$59.7\,    $&$0.9  $&$6.4  $&$4.4  $&$4.6  $&$1.5  $&$0.1  $&$0.5  $&$4.0  $&$0.91 $&$2.6$ \\ 
7 &$21.0\,    $&$0.7  $&$5.9  $&$4.1  $&$4.3  $&$1.3  $&$0.1  $&$0.6  $&$3.7  $&$0.92 $&$1.5$ \\ 
\hline

\multicolumn{12}{c}{ } \\
\multicolumn{12}{c}{\normalsize 2-Jet Cross Section ${\rm  \frac{d\sigma_{2-jet}}{dQ^2}}$ }\\
\multicolumn{12}{c}{ } \\
\hline
 & & &  & total & total &
\multicolumn{4}{|c||}{\underline{\hspace*{0.1cm}single contributions to 
                       correlated uncertainty\hspace*{0.1cm}} }
   &                                           hadronisation & hadronisation\\
bin   & cross        & statistical  & total        & uncorrelated & correlated & model  & electron     & electron   & hadronic    & correction    & correction    \\
      & section      & uncert.      & uncert.      & uncertainty  & uncert.    & uncert.& energy scale & polar angle&  energy scale &  factor    & uncertainty \\
    & $\rm [pb/GeV^{2}]$ & [\%]    & [\%]         & [\%]         & [\%]       &   [\%] & [\%]         & [\%]       & [\%] &       & [\%]\\
\hline
1 &$83.6\,    $&$1.5  $&$6.8  $&$4.8  $&$4.8  $&$1.2  $&$0.2  $&$0.2  $&$4.4  $&$0.91 $&$2.8$ \\ 
2 &$56.4\,    $&$1.4  $&$8.4  $&$5.9  $&$6.0  $&$3.6  $&$0.1  $&$0.5  $&$4.5  $&$0.93 $&$2.9$ \\ 
3 &$33.9\,    $&$1.3  $&$7.4  $&$5.2  $&$5.2  $&$1.5  $&$0.1  $&$0.7  $&$4.7  $&$0.93 $&$1.8$ \\ 
4 &$22.5\,    $&$1.6  $&$7.7  $&$5.5  $&$5.4  $&$2.0  $&$0.1  $&$0.7  $&$4.8  $&$0.94 $&$1.8$ \\ 
5 &$14.4\,    $&$1.4  $&$7.2  $&$5.1  $&$5.1  $&$1.5  $&$0.2  $&$0.5  $&$4.6  $&$0.94 $&$2.0$ \\ 
6 &$9.70\,     $&$1.9  $&$7.3  $&$5.2  $&$5.1  $&$1.5  $&$0.1  $&$0.6  $&$4.6  $&$0.94 $&$1.9$ \\ 
7 &$3.72\,     $&$1.4  $&$7.2  $&$5.1  $&$5.1  $&$1.2  $&$0.2  $&$0.6  $&$4.7  $&$0.95 $&$1.9$ \\ 
\hline

\multicolumn{12}{c}{ } \\
\multicolumn{12}{c}{\normalsize 3-Jet Cross Section ${\rm \frac{d\sigma_{3-jet}}{dQ^2}}$ }\\
\multicolumn{12}{c}{ } \\
\hline
 & & &  & total & total &
\multicolumn{4}{|c||}{\underline{\hspace*{0.1cm}single contributions to 
                       correlated uncertainty\hspace*{0.1cm}} }
   &                                           hadronisation & hadronisation\\
bin   & cross        & statistical  & total        & uncorrelated & correlated & model  & electron     & electron   & hadronic    & correction    & correction    \\
      & section      & uncert.      & uncert.      & uncertainty  & uncert.    & uncert.& energy scale & polar angle&  energy scale &  factor    & uncertainty \\
    & $\rm [pb/GeV^{2}]$ & [\%]    & [\%]         & [\%]         & [\%]       &   [\%] & [\%]         & [\%]       & [\%]&       & [\%]\\

\hline
I   &$10.5\,   $&$2.6  $&$10.8 $&$7.7  $&$7.5  $&$4.6  $&$0.4  $&$0.4  $&$5.6  $&$0.81 $&$6.3$ \\ 
II  &$4.56\,    $&$2.7  $&$10.6 $&$7.7  $&$7.4  $&$4.0  $&$0.7  $&$0.8  $&$5.9  $&$0.82 $&$5.5$ \\ 
III &$2.05\,    $&$2.9  $&$10.8 $&$7.9  $&$7.5  $&$4.1  $&$0.7  $&$0.6  $&$6.0  $&$0.80 $&$7.2$ \\ 
IV  &$0.560\,    $&$3.8  $&$11.7 $&$8.6  $&$7.9  $&$4.5  $&$1.0  $&$0.6  $&$6.2  $&$0.81 $&$6.1$ \\ 
\hline

\end{tabular} 

\normalfont
 \vspace{\baselineskip} 
\caption{\label{tab::jet_Q2} 
Inclusive jet, 2-jet and 3-jet cross sections in NC DIS measured 
as a function of $Q^2$. The measurements refer to the phase-space 
given in table \ref{tab:selection}. 
In the columns 3 to 6 are shown the statistical 
uncertainty, 
the total experimental uncertainty, the total uncorrelated uncertainty including the statistical 
one and the total bin-to-bin correlated uncertainty calculated as the quadratic sum of the 
following five components: the model dependence, the electron and the hadronic energy scales, 
the electron polar angle  and the luminosity measurement uncertainties.
The contributions to correlated uncertainties are listed in columns 7 to 10.
The sharing of the uncertainties between correlated and uncorrelated sources is 
described in detail in section \ref{sec:sys}. The hadronisation correction factors applied to 
the NLO predictions and their uncertainties are shown in columns 11 and 12. The bin nomenclature of 
column 1 is defined in table \ref{tab:bins}.
} 
\end{table}

\begin{table}[htp]
\centering
\tiny \sf
\renewcommand{\arraystretch}{1.50}
\begin{tabular}{|c || r || r | r | r | r || r | r | r | r || r | r|} 

\multicolumn{12}{c}{ } \\
\multicolumn{12}{c}{\normalsize Inclusive Jet Cross Section ${\rm \frac{d\sigma_{jet}}{dP_T}}$ }\\
\multicolumn{12}{c}{ } \\
\hline
 & & &  & total & total &
\multicolumn{4}{|c||}{\underline{\hspace*{0.1cm}single contributions to 
                       correlated uncertainty\hspace*{0.1cm}} }
   &                                           hadronisation & hadronisation\\
bin   & cross        & statistical  & total        & uncorrelated & correlated & model  & electron     & electron   & hadronic    & correction    & correction    \\
      & section      & uncert.      & uncert.      & uncertainty  & uncert.    & uncert.& energy scale & polar angle&  energy scale &  factor    & uncertainty \\
    & $\rm [pb/GeV]$ & [\%]    & [\%]         & [\%]         & [\%]       &   [\%] & [\%]         & [\%]       & [\%] &       & [\%]\\
\hline
 a&$1250\,  $&$0.3  $&$6.8  $&$4.7  $&$4.9  $&$2.1  $&$0.2  $&$0.4  $&$4.1  $&$0.89 $&$4.4$ \\ 
 b&$174\,   $&$0.7  $&$6.6  $&$4.5  $&$4.7  $&$2.1  $&$0.1  $&$0.5  $&$3.9  $&$0.93 $&$2.8$ \\ 
 c&$38.3\,    $&$1.5  $&$9.8  $&$6.9  $&$6.9  $&$3.4  $&$0.3  $&$0.6  $&$5.8  $&$0.95 $&$2.2$ \\ 
 d&$1.53\,     $&$2.2  $&$15.3 $&$10.9 $&$10.8 $&$8.1  $&$0.6  $&$0.8  $&$6.9  $&$0.95 $&$1.5$ \\ 
\hline

\multicolumn{12}{c}{ } \\
\multicolumn{12}{c}{\normalsize 2-Jet Cross Section ${\rm \frac{d\sigma_{2-jet}}{d\Ptav}}$ }\\
\multicolumn{12}{c}{ } \\
\hline
 & & &  & total & total &
\multicolumn{4}{|c||}{\underline{\hspace*{0.1cm}single contributions to 
                       correlated uncertainty\hspace*{0.1cm}} }
   &                                           hadronisation & hadronisation\\
bin   & cross        & statistical  & total        & uncorrelated & correlated & model  & electron     & electron   & hadronic    & correction    & correction    \\
      & section      & uncert.      & uncert.      & uncertainty  & uncert.    & uncert.& energy scale & polar angle&  energy scale &  factor    & uncertainty \\
    & $\rm [pb/GeV]$ & [\%]    & [\%]         & [\%]         & [\%]       &   [\%] & [\%]         & [\%]       & [\%] &       & [\%]\\
\hline
 a&$120\,   $&$0.8  $&$6.8  $&$4.7  $&$4.9  $&$2.1  $&$0.2  $&$0.4  $&$4.1  $&$0.92 $&$2.1$ \\ 
 b&$72.5\,    $&$0.9  $&$7.7  $&$5.4  $&$5.5  $&$2.3  $&$0.1  $&$0.5  $&$4.7  $&$0.94 $&$2.3$ \\ 
 c&$15.5\,    $&$2.0  $&$11.1 $&$7.9  $&$7.8  $&$4.9  $&$0.4  $&$0.6  $&$5.9  $&$0.95 $&$2.3$ \\ 
 d&$0.620\,     $&$3.0  $&$17.8 $&$12.7 $&$12.5 $&$9.7  $&$0.6  $&$0.8  $&$7.7  $&$0.96 $&$1.8$ \\
\hline

\multicolumn{12}{c}{ } \\
\multicolumn{12}{c}{\normalsize 3-Jet Cross Section ${\rm \frac{d\sigma_{3-jet}}{d\Ptav}}$ }\\
\multicolumn{12}{c}{ } \\
\hline
 & & &  & total & total &
\multicolumn{4}{|c||}{\underline{\hspace*{0.1cm}single contributions to 
                       correlated uncertainty\hspace*{0.1cm}} }
   &                                           hadronisation & hadronisation\\
bin   & cross        & statistical  & total        & uncorrelated & correlated & model  & electron     & electron   & hadronic    & correction    & correction    \\
      & section      & uncert.      & uncert.      & uncertainty  & uncert.    & uncert.& energy scale & polar angle&  energy scale &  factor    & uncertainty \\
    & $\rm [pb/GeV]$ & [\%]    & [\%]         & [\%]         & [\%]       &   [\%] & [\%]         & [\%]       & [\%] &       & [\%]\\
\hline
 a&$11.7\,    $&$2.5  $&$9.2  $&$6.6  $&$6.3  $&$2.4  $&$0.5  $&$0.5  $&$5.6  $&$0.77 $&$9.5$ \\ 
 b&$16.1\,    $&$2.1  $&$9.0  $&$6.4  $&$6.3  $&$2.5  $&$0.6  $&$0.6  $&$5.5  $&$0.81 $&$6.5$ \\ 
 c&$4.29\,     $&$3.9  $&$12.7 $&$9.4  $&$8.6  $&$5.3  $&$0.8  $&$0.7  $&$6.6  $&$0.84 $&$3.6$ \\ 
 d&$0.192\,     $&$5.6  $&$20.0 $&$14.7 $&$13.6 $&$10.6 $&$1.0  $&$0.9  $&$8.4  $&$0.87 $&$4.1$ \\ 
\hline

\end{tabular} 
\normalfont
 \vspace{\baselineskip} 
\caption{\label{tab::jet_Pt} 
Inclusive jet, 2-jet and 3-jet cross sections in NC DIS measured 
as a function of $P_T$ for inclusive jet and $\Ptav$ of the two
leading jets for 2-jet and 3-jet cross sections
together with their relative errors and hadronisation correction
factors. Other details are given in the caption to table \ref{tab::jet_Q2}.
The bin nomenclature is defined in table \ref{tab:bins}.
} 
\end{table}


\begin{table}[htp]
\centering
\tiny \sf
\renewcommand{\arraystretch}{1.50}
\begin{tabular}{|c || r || r | r | r | r || r | r | r | r || r | r|} 

\multicolumn{12}{c}{ } \\
\multicolumn{12}{c}{\normalsize Inclusive Jet Cross Section ${\rm \frac{d^2\sigma_{jet}}{dQ^2dP_T}}$ }\\
\multicolumn{12}{c}{ } \\
\hline
 & & &  & total & total &
\multicolumn{4}{|c||}{\underline{\hspace*{0.1cm}single contributions to 
                       correlated uncertainty\hspace*{0.1cm}} }
   &                                           hadronisation & hadronisation\\
bin   & cross        & statistical  & total        & uncorrelated & correlated & model  & electron     & electron   & hadronic    & correction    & correction    \\
      & section      & uncert.      & uncert.      & uncertainty  & uncert.    & uncert.& energy scale & polar angle&  energy scale &  factor    & uncertainty \\
    & $\rm [pb/GeV^{3}]$ & [\%]    & [\%]         & [\%]         & [\%]       &   [\%] & [\%]         & [\%]       & [\%] &       & [\%]\\
\hline
1 a&$109\,       $&$0.7  $&$7.1  $&$4.9  $&$5.1  $&$2.1  $&$0.4  $&$0.5  $&$4.3  $&$0.87 $&$6.7$ \\ 
1 b&$12.9\,       $&$1.8  $&$6.1  $&$4.4  $&$4.2  $&$2.7  $&$0.1  $&$0.1  $&$2.9  $&$0.91 $&$3.2$ \\ 
1 c&$2.77\,       $&$3.8  $&$10.1 $&$7.6  $&$6.7  $&$3.4  $&$0.2  $&$0.1  $&$5.6  $&$0.95 $&$3.4$ \\ 
1 d&$0.104\,      $&$6.4  $&$15.0 $&$11.4 $&$9.6  $&$6.6  $&$1.0  $&$1.0  $&$6.7  $&$0.95 $&$3.4$ \\ 
\hline 
2 a&$72.1\,       $&$0.7  $&$7.6  $&$5.3  $&$5.5  $&$2.9  $&$0.3  $&$0.4  $&$4.4  $&$0.88 $&$6.1$ \\ 
2 b&$9.02\,       $&$1.7  $&$6.6  $&$4.7  $&$4.6  $&$2.9  $&$0.1  $&$0.5  $&$3.3  $&$0.92 $&$2.9$ \\ 
2 c&$1.93\,       $&$3.4  $&$15.8 $&$11.4 $&$11.0 $&$9.4  $&$0.3  $&$0.8  $&$5.4  $&$0.95 $&$3.1$ \\ 
2 d&$0.0729\,      $&$5.8  $&$24.6 $&$17.8 $&$17.0 $&$15.5 $&$0.1  $&$0.6  $&$6.8  $&$0.95 $&$3.1$ \\ 
\hline 
3 a&$42.8\,       $&$0.6  $&$7.6  $&$5.3  $&$5.4  $&$2.7  $&$0.2  $&$0.6  $&$4.4  $&$0.89 $&$5.2$ \\ 
3 b&$5.45\,       $&$1.6  $&$7.0  $&$5.0  $&$4.9  $&$2.8  $&$0.2  $&$0.7  $&$3.7  $&$0.92 $&$2.9$ \\ 
3 c&$1.12\,       $&$3.3  $&$12.2 $&$8.9  $&$8.4  $&$6.0  $&$0.2  $&$0.8  $&$5.6  $&$0.95 $&$2.8$ \\ 
3 d&$0.0484\,      $&$5.5  $&$20.5 $&$15.0 $&$14.0 $&$12.0 $&$0.7  $&$1.0  $&$7.0  $&$0.95 $&$2.8$ \\ 
\hline 
4 a&$26.7\,       $&$0.8  $&$7.3  $&$5.1  $&$5.2  $&$2.7  $&$0.1  $&$0.6  $&$4.2  $&$0.90 $&$4.3$ \\ 
4 b&$3.68\,       $&$1.9  $&$7.6  $&$5.4  $&$5.3  $&$2.7  $&$0.1  $&$0.7  $&$4.2  $&$0.93 $&$2.7$ \\ 
4 c&$0.783\,      $&$3.9  $&$12.6 $&$9.2  $&$8.5  $&$6.1  $&$0.4  $&$0.8  $&$5.7  $&$0.94 $&$4.1$ \\ 
4 d&$0.0326\,      $&$6.5  $&$17.2 $&$13.0 $&$11.3 $&$8.9  $&$0.5  $&$0.8  $&$6.8  $&$0.95 $&$4.1$ \\ 
\hline 
5 a&$16.4\,       $&$0.7  $&$7.0  $&$4.8  $&$5.0  $&$2.5  $&$0.3  $&$0.5  $&$4.1  $&$0.90 $&$3.7$ \\ 
5 b&$2.21\,        $&$1.8  $&$7.5  $&$5.4  $&$5.3  $&$2.5  $&$0.1  $&$0.4  $&$4.4  $&$0.94 $&$3.0$ \\ 
5 c&$0.508\,      $&$3.5  $&$10.2 $&$7.6  $&$6.9  $&$2.8  $&$0.3  $&$0.6  $&$6.1  $&$0.95 $&$3.0$ \\ 
5 d&$0.0198\,      $&$5.6  $&$15.0 $&$11.2 $&$9.9  $&$7.1  $&$0.4  $&$0.6  $&$6.7  $&$0.96 $&$2.9$ \\ 
\hline 
6 a&$9.72\,       $&$1.0  $&$7.3  $&$5.1  $&$5.2  $&$3.2  $&$0.1  $&$0.6  $&$3.8  $&$0.91 $&$2.6$ \\ 
6 b&$1.65\,       $&$2.2  $&$8.6  $&$6.2  $&$5.9  $&$3.3  $&$0.2  $&$0.5  $&$4.6  $&$0.95 $&$2.6$ \\ 
6 c&$0.383\,      $&$4.5  $&$11.1 $&$8.4  $&$7.3  $&$3.7  $&$0.2  $&$0.5  $&$6.1  $&$0.96 $&$5.2$ \\ 
6 d&$0.0120\,      $&$7.5  $&$15.0 $&$11.8 $&$9.3  $&$5.4  $&$0.7  $&$0.6  $&$7.3  $&$0.94 $&$2.8$ \\ 
\hline 
7 a&$3.37\,       $&$0.8  $&$6.5  $&$4.5  $&$4.7  $&$2.8  $&$0.1  $&$0.7  $&$3.4  $&$0.91 $&$1.5$ \\ 
7 b&$0.614\,      $&$1.7  $&$9.5  $&$6.7  $&$6.7  $&$4.2  $&$0.2  $&$0.6  $&$5.0  $&$0.96 $&$1.7$ \\ 
7 c&$0.146\,      $&$3.3  $&$10.6 $&$7.8  $&$7.2  $&$2.8  $&$0.4  $&$0.6  $&$6.4  $&$0.94 $&$2.7$ \\ 
7 d&$0.00677\,    $&$5.3  $&$12.0 $&$9.2  $&$7.7  $&$2.5  $&$0.6  $&$0.6  $&$7.1  $&$0.96 $&$2.6$ \\ 
\hline

\end{tabular}
\normalfont
 \vspace{\baselineskip}
\caption{\label{tab::Ijet_Q2PT}
Double differential inclusive jet cross sections as
a function of $Q^2$ and $P_T$ together with their 
relative errors and hadronisation correction
factors. Other details are given in the caption to table \ref{tab::jet_Q2}.
The bin nomenclature is defined in table \ref{tab:bins}.}
\end{table}

\begin{table}[htp]
\centering
\tiny \sf
\renewcommand{\arraystretch}{1.50}
\begin{tabular}{|c || r || r | r | r | r || r | r | r | r || r | r|} 

\multicolumn{12}{c}{ } \\
\multicolumn{12}{c}{\normalsize 2-Jet Cross Section ${\rm \frac{d^2\sigma_{2-jet}}{dQ^2d\Ptav}}$ }\\
\multicolumn{12}{c}{ } \\
\hline
 & & &  & total & total &
\multicolumn{4}{|c||}{\underline{\hspace*{0.1cm}single contributions to 
                       correlated uncertainty\hspace*{0.1cm}} }
   &                                           hadronisation & hadronisation\\
bin   & cross        & statistical  & total        & uncorrelated & correlated & model  & electron     & electron   & hadronic    & correction    & correction    \\
      & section      & uncert.      & uncert.      & uncertainty  & uncert.    & uncert.& energy scale & polar angle&  energy scale &  factor    & uncertainty \\
    & $\rm [pb/GeV^{3}]$ & [\%]    & [\%]         & [\%]         & [\%]       &   [\%] & [\%]         & [\%]       & [\%] &       & [\%]\\
\hline
1 a&$10.1\,       $&$2.0  $&$6.7  $&$4.8  $&$4.6  $&$1.9  $&$0.6  $&$0.5  $&$3.9  $&$0.91 $&$3.3$ \\ 
1 b&$5.06\,       $&$2.6  $&$7.6  $&$5.6  $&$5.1  $&$2.2  $&$0.1  $&$0.2  $&$4.4  $&$0.91 $&$2.8$ \\ 
1 c&$1.08\,       $&$5.5  $&$11.5 $&$8.9  $&$7.2  $&$4.6  $&$0.4  $&$0.2  $&$5.4  $&$0.95 $&$2.8$ \\ 
1 d&$0.0435\,     $&$8.3  $&$16.7 $&$13.1 $&$10.3 $&$6.3  $&$1.0  $&$0.9  $&$7.9  $&$0.96 $&$2.7$ \\ 
\hline 
2 a&$6.55\,       $&$1.8  $&$6.7  $&$4.8  $&$4.7  $&$1.7  $&$0.3  $&$0.3  $&$4.1  $&$0.91 $&$3.4$ \\ 
2 b&$3.56\,       $&$2.4  $&$8.2  $&$6.0  $&$5.7  $&$3.1  $&$0.1  $&$0.6  $&$4.5  $&$0.93 $&$3.0$ \\ 
2 c&$0.767\,      $&$5.0  $&$18.8 $&$13.7 $&$12.9 $&$11.4 $&$0.2  $&$0.8  $&$5.6  $&$0.95 $&$2.4$ \\ 
2 d&$0.0289\,     $&$7.8  $&$27.5 $&$20.2 $&$18.7 $&$17.1 $&$0.2  $&$0.8  $&$7.3  $&$0.94 $&$3.0$ \\ 
\hline 
3 a&$3.78\,       $&$1.8  $&$7.4  $&$5.2  $&$5.1  $&$2.1  $&$0.2  $&$0.7  $&$4.4  $&$0.92 $&$1.7$ \\ 
3 b&$2.28\,       $&$2.2  $&$7.9  $&$5.7  $&$5.4  $&$2.3  $&$0.2  $&$0.7  $&$4.6  $&$0.94 $&$2.8$ \\ 
3 c&$0.441\,      $&$4.9  $&$15.1 $&$11.2 $&$10.2 $&$7.9  $&$0.3  $&$0.7  $&$6.1  $&$0.94 $&$2.8$ \\ 
3 d&$0.0201\,     $&$7.3  $&$22.7 $&$16.8 $&$15.2 $&$13.1 $&$0.7  $&$1.0  $&$7.5  $&$0.96 $&$3.0$ \\ 
\hline 
4 a&$2.45\,       $&$2.3  $&$7.3  $&$5.3  $&$5.0  $&$1.8  $&$0.2  $&$0.7  $&$4.4  $&$0.94 $&$1.5$ \\ 
4 b&$1.52\,       $&$2.7  $&$8.1  $&$5.9  $&$5.5  $&$2.0  $&$0.1  $&$0.8  $&$4.8  $&$0.94 $&$2.2$ \\ 
4 c&$0.327\,      $&$5.7  $&$13.5 $&$10.3 $&$8.8  $&$6.3  $&$0.4  $&$0.9  $&$5.8  $&$0.92 $&$1.8$ \\ 
4 d&$0.0135\,     $&$8.7  $&$20.6 $&$15.8 $&$13.3 $&$10.8 $&$0.6  $&$0.8  $&$7.5  $&$0.97 $&$3.0$ \\ 
\hline 
5 a&$1.64\,       $&$2.0  $&$6.8  $&$4.9  $&$4.7  $&$1.8  $&$0.4  $&$0.4  $&$4.1  $&$0.93 $&$2.0$ \\ 
5 b&$0.927\,      $&$2.5  $&$8.1  $&$5.9  $&$5.6  $&$2.2  $&$0.1  $&$0.5  $&$4.9  $&$0.95 $&$2.2$ \\ 
5 c&$0.213\,      $&$5.1  $&$10.5 $&$8.2  $&$6.6  $&$2.5  $&$0.3  $&$0.6  $&$5.8  $&$0.96 $&$2.2$ \\ 
5 d&$0.00763\,    $&$7.8  $&$20.1 $&$15.2 $&$13.1 $&$10.3 $&$0.5  $&$0.6  $&$8.0  $&$0.96 $&$2.8$ \\ 
\hline 
6 a&$0.968\,      $&$2.7  $&$7.2  $&$5.3  $&$4.8  $&$2.3  $&$0.1  $&$0.6  $&$4.0  $&$0.93 $&$1.8$ \\ 
6 b&$0.753\,      $&$3.1  $&$8.8  $&$6.5  $&$5.9  $&$2.6  $&$0.2  $&$0.5  $&$5.0  $&$0.95 $&$1.9$ \\ 
6 c&$0.138\,      $&$6.3  $&$11.5 $&$9.2  $&$6.9  $&$2.8  $&$0.4  $&$0.5  $&$6.0  $&$0.95 $&$4.5$ \\ 
6 d&$0.00521\,    $&$9.9  $&$18.1 $&$14.5 $&$10.8 $&$7.0  $&$0.7  $&$0.7  $&$8.0  $&$0.97 $&$4.4$ \\ 
\hline 
7 a&$0.379\,      $&$2.1  $&$6.7  $&$4.9  $&$4.7  $&$1.7  $&$0.1  $&$0.6  $&$4.0  $&$0.94 $&$1.8$ \\ 
7 b&$0.266\,      $&$2.4  $&$8.2  $&$5.9  $&$5.6  $&$1.8  $&$0.4  $&$0.6  $&$5.1  $&$0.96 $&$1.8$ \\ 
7 c&$0.0648\,     $&$4.8  $&$11.2 $&$8.6  $&$7.3  $&$1.8  $&$0.6  $&$0.6  $&$6.8  $&$0.96 $&$2.7$ \\ 
7 d&$0.00265\,    $&$7.0  $&$15.7 $&$12.1 $&$10.0 $&$5.9  $&$0.5  $&$0.5  $&$7.9  $&$0.95 $&$1.8$ \\ 
\hline 

\end{tabular} 
\normalfont
 \vspace{\baselineskip} 
\caption{\label{tab::Dijet_Q2PT} 
Double differential 2-jet cross sections as 
a function of $Q^2$ and $\Ptav$ of two leading jets together with their relative 
errors and hadronisation correction 
factors. Other details are given in the caption to table \ref{tab::jet_Q2}. 
The bin nomenclature is defined in table \ref{tab:bins}.} 
\end{table}

\begin{table}[htp]
\centering
\tiny \sf
\renewcommand{\arraystretch}{1.50}
\begin{tabular}{|c || r || r | r | r | r || r | r | r | r || r | r|} 

\multicolumn{12}{c}{ } \\
\multicolumn{12}{c}{\normalsize 2-Jet Cross Section ${\rm \frac{d^2\sigma_{2-jet}}{dQ^2d\xi}}$ }\\
\multicolumn{12}{c}{ } \\
\hline
 & & &  & total & total &
\multicolumn{4}{|c||}{\underline{\hspace*{0.1cm}single contributions to 
                       correlated uncertainty\hspace*{0.1cm}} }
   &                                           hadronisation & hadronisation\\
bin   & cross        & statistical  & total        & uncorrelated & correlated & model  & electron     & electron   & hadronic    & correction    & correction    \\
      & section      & uncert.      & uncert.      & uncertainty  & uncert.    & uncert.& energy scale & polar angle&  energy scale &  factor    & uncertainty \\
    & $\rm [pb/GeV^{2}]$ & [\%]    & [\%]         & [\%]         & [\%]       &   [\%] & [\%]         & [\%]       & [\%] &       & [\%]\\
\hline
1 A&$608\,     $&$9.7  $&$35.3 $&$25.9 $&$24.1 $&$23.6 $&$1.2  $&$0.6  $&$4.3  $&$0.94 $&$10.2$ \\ 
1 B&$3450\,    $&$3.4  $&$10.0 $&$7.4  $&$6.7  $&$5.5  $&$0.5  $&$0.3  $&$3.5  $&$0.96 $&$7.0$ \\ 
1 C&$3060\,    $&$2.1  $&$7.2  $&$5.2  $&$5.0  $&$2.8  $&$0.8  $&$1.0  $&$3.6  $&$0.92 $&$3.3$ \\ 
1 D&$687\,     $&$3.4  $&$8.4  $&$6.4  $&$5.4  $&$2.7  $&$0.4  $&$1.4  $&$4.2  $&$0.87 $&$3.3$ \\ 
1 E&$76.6\,     $&$7.2  $&$15.1 $&$11.8 $&$9.4  $&$7.4  $&$0.6  $&$1.1  $&$5.4  $&$0.85 $&$3.4$ \\ 
1 F&$4.40\,     $&$15.8 $&$35.0 $&$27.2 $&$22.1 $&$20.7 $&$1.3  $&$0.9  $&$7.2  $&$0.92 $&$3.1$ \\ 
\hline 
2 A&$723\,     $&$9.7  $&$12.4 $&$11.1 $&$5.6  $&$4.4  $&$1.8  $&$0.8  $&$2.3  $&$0.97 $&$7.6$ \\ 
2 B&$2470\,    $&$3.3  $&$6.9  $&$5.3  $&$4.4  $&$2.2  $&$0.7  $&$0.6  $&$3.4  $&$0.97 $&$7.0$ \\ 
2 C&$2060\,    $&$1.8  $&$6.8  $&$4.9  $&$4.7  $&$2.2  $&$0.3  $&$0.4  $&$3.9  $&$0.93 $&$3.6$ \\ 
2 D&$441\,     $&$3.0  $&$10.0 $&$7.3  $&$6.8  $&$5.1  $&$0.3  $&$0.5  $&$4.2  $&$0.89 $&$1.7$ \\ 
2 E&$50.7\,     $&$6.3  $&$19.4 $&$14.4 $&$13.0 $&$11.5 $&$0.9  $&$0.5  $&$5.8  $&$0.87 $&$1.7$ \\ 
2 F&$2.31\,     $&$15.2 $&$42.2 $&$31.7 $&$27.9 $&$26.9 $&$1.7  $&$0.4  $&$7.2  $&$0.87 $&$2.0$ \\ 
\hline 
3 A&$363\,     $&$10.4 $&$13.6 $&$12.1 $&$6.3  $&$1.5  $&$0.6  $&$1.0  $&$5.8  $&$1.02 $&$5.2$ \\ 
3 B&$1430\,    $&$3.3  $&$6.9  $&$5.3  $&$4.4  $&$1.6  $&$0.9  $&$0.7  $&$3.7  $&$0.97 $&$4.5$ \\ 
3 C&$1250\,    $&$1.8  $&$6.7  $&$4.8  $&$4.7  $&$1.7  $&$0.2  $&$0.7  $&$4.1  $&$0.93 $&$1.9$ \\ 
3 D&$261\,     $&$3.0  $&$7.4  $&$5.5  $&$4.9  $&$1.2  $&$0.2  $&$0.7  $&$4.4  $&$0.90 $&$2.1$ \\ 
3 E&$38.2\,     $&$5.6  $&$18.6 $&$13.7 $&$12.6 $&$10.8 $&$1.4  $&$0.9  $&$6.1  $&$0.88 $&$2.1$ \\ 
3 F&$1.21\,     $&$16.0 $&$38.0 $&$29.2 $&$24.3 $&$23.0 $&$2.3  $&$1.2  $&$7.2  $&$0.87 $&$4.0$ \\ 
\hline 
4 A&$231\,     $&$13.9 $&$15.0 $&$14.4 $&$4.2  $&$1.8  $&$0.8  $&$0.6  $&$3.4  $&$1.09 $&$3.0$ \\ 
4 B&$1020\,    $&$4.0  $&$7.9  $&$6.2  $&$5.0  $&$1.6  $&$1.0  $&$0.6  $&$4.3  $&$0.97 $&$5.1$ \\ 
4 C&$782\,     $&$2.2  $&$6.7  $&$4.9  $&$4.6  $&$1.0  $&$0.3  $&$0.7  $&$4.1  $&$0.95 $&$1.9$ \\ 
4 D&$189\,     $&$3.5  $&$7.7  $&$5.9  $&$5.0  $&$1.1  $&$0.3  $&$0.8  $&$4.5  $&$0.90 $&$2.0$ \\ 
4 E&$26.3\,     $&$6.7  $&$17.2 $&$13.0 $&$11.2 $&$9.2  $&$1.3  $&$0.8  $&$6.0  $&$0.87 $&$1.5$ \\ 
4 F&$0.946\,    $&$17.4 $&$26.7 $&$22.5 $&$14.3 $&$13.0 $&$1.5  $&$0.7  $&$5.5  $&$0.90 $&$1.4$ \\ 
\hline 
5 A&$82\,     $&$14.3 $&$15.4 $&$14.8 $&$4.3  $&$2.2  $&$0.6  $&$0.5  $&$3.3  $&$1.04 $&$4.3$ \\ 
5 B&$602\,     $&$3.7  $&$7.0  $&$5.5  $&$4.3  $&$2.1  $&$1.1  $&$0.4  $&$3.2  $&$0.98 $&$3.8$ \\ 
5 C&$523\,     $&$1.9  $&$6.8  $&$4.9  $&$4.7  $&$1.6  $&$0.4  $&$0.5  $&$4.1  $&$0.95 $&$2.4$ \\ 
5 D&$120\,     $&$3.1  $&$7.9  $&$5.9  $&$5.2  $&$1.7  $&$0.4  $&$0.6  $&$4.6  $&$0.90 $&$1.0$ \\ 
5 E&$16.6\,     $&$5.9  $&$11.2 $&$8.9  $&$6.9  $&$2.9  $&$1.0  $&$0.6  $&$5.9  $&$0.87 $&$1.1$ \\ 
5 F&$0.700\,    $&$14.3 $&$29.5 $&$23.1 $&$18.3 $&$16.4 $&$1.9  $&$0.9  $&$7.7  $&$0.91 $&$1.7$ \\ 
\hline 
6 A&$33.9\,     $&$20.9 $&$21.6 $&$21.2 $&$4.3  $&$2.1  $&$1.7  $&$0.2  $&$2.9  $&$0.98 $&$6.8$ \\ 
6 B&$431\,     $&$5.0  $&$7.3  $&$6.2  $&$3.9  $&$0.9  $&$0.7  $&$0.6  $&$3.4  $&$0.99 $&$2.3$ \\ 
6 C&$335\,     $&$2.5  $&$7.1  $&$5.3  $&$4.8  $&$2.2  $&$0.4  $&$0.5  $&$4.0  $&$0.95 $&$2.9$ \\ 
6 D&$93.8\,     $&$4.0  $&$8.7  $&$6.7  $&$5.6  $&$2.2  $&$0.3  $&$0.5  $&$4.8  $&$0.91 $&$0.9$ \\ 
6 E&$9.52\,     $&$7.9  $&$12.6 $&$10.5 $&$7.1  $&$3.0  $&$1.5  $&$0.7  $&$6.0  $&$0.89 $&$1.0$ \\ 
6 F&$0.427\,    $&$18.9 $&$27.7 $&$23.7 $&$14.4 $&$12.3 $&$1.8  $&$0.5  $&$6.9  $&$0.91 $&$2.2$ \\ 
\hline 
7 A&$4.83\,     $&$44.7 $&$45.5 $&$45.1 $&$6.3  $&$4.8  $&$0.8  $&$0.8  $&$3.5  $&$1.17 $&$5.2$ \\ 
7 B&$122\,     $&$5.2  $&$9.6  $&$7.7  $&$5.7  $&$3.7  $&$0.9  $&$0.3  $&$4.0  $&$0.98 $&$1.9$ \\ 
7 C&$131\,     $&$2.0  $&$7.2  $&$5.1  $&$5.0  $&$2.9  $&$0.1  $&$0.4  $&$3.8  $&$0.96 $&$1.4$ \\ 
7 D&$36.5\,     $&$2.7  $&$8.8  $&$6.5  $&$5.9  $&$3.1  $&$0.6  $&$0.5  $&$4.8  $&$0.94 $&$1.4$ \\ 
7 E&$5.81\,     $&$4.9  $&$11.6 $&$8.9  $&$7.5  $&$3.1  $&$1.5  $&$0.4  $&$6.4  $&$0.90 $&$1.5$ \\ 
7 F&$0.237\,    $&$11.8 $&$22.1 $&$17.7 $&$13.2 $&$10.3 $&$2.3  $&$0.4  $&$7.8  $&$0.89 $&$2.7$ \\ 
\hline

\end{tabular} 
\normalfont
 \vspace{\baselineskip} 
\caption{\label{tab::Dijet_Q2Ksi} 
Double differential 2-jet cross sections as 
a function of $Q^2$ and $\xi$ together with their relative errors 
and hadronisation correction 
factors. Other details are given in the caption to table \ref{tab::jet_Q2}. 
The bin nomenclature is defined in table \ref{tab:bins}.} 
\end{table}

\begin{table}[htp]
\centering
\tiny \sf
\renewcommand{\arraystretch}{1.50}
\begin{tabular}{|c || r || r | r | r | r || r | r | r | r || r | r|} 

\multicolumn{12}{c}{ } \\
\multicolumn{12}{c}{\normalsize 3-Jet Cross Section ${\rm \frac{d^2\sigma_{3-jet}}{dQ^2d\Ptav}}$ }\\
\multicolumn{12}{c}{ } \\
\hline
 & & &  & total & total &
\multicolumn{4}{|c||}{\underline{\hspace*{0.1cm}single contributions to 
                       correlated uncertainty\hspace*{0.1cm}} }
   &                                           hadronisation & hadronisation\\
bin   & cross        & statistical  & total        & uncorrelated & correlated & model  & electron     & electron   & hadronic    & correction    & correction    \\
      & section      & uncert.      & uncert.      & uncertainty  & uncert.    & uncert.& energy scale & polar angle&  energy scale &  factor    & uncertainty \\
    & $\rm [pb/GeV^{3}]$ & [\%]    & [\%]         & [\%]         & [\%]       &   [\%] & [\%]         & [\%]       & [\%] &       & [\%]\\
\hline
I a&$0.780\,      $&$4.3  $&$10.0 $&$7.6  $&$6.5  $&$3.2  $&$0.3  $&$0.2  $&$5.4  $&$0.78 $&$11.4$ \\ 
I b&$0.951\,      $&$3.8  $&$9.6  $&$7.2  $&$6.4  $&$3.4  $&$0.3  $&$0.4  $&$5.1  $&$0.81 $&$6.0$ \\ 
I c&$0.240\,      $&$7.2  $&$20.4 $&$15.2 $&$13.5 $&$11.8 $&$0.7  $&$0.6  $&$6.3  $&$0.82 $&$5.8$ \\ 
I d&$0.0102\,     $&$10.9 $&$24.3 $&$18.8 $&$15.4 $&$12.6 $&$0.8  $&$1.1  $&$8.5  $&$0.88 $&$5.5$ \\ 
\hline 
II a&$0.321\,     $&$4.5  $&$10.5 $&$8.0  $&$6.8  $&$2.9  $&$0.6  $&$0.9  $&$5.8  $&$0.78 $&$8.4$ \\ 
II b&$0.409\,     $&$3.9  $&$10.0 $&$7.5  $&$6.6  $&$3.1  $&$0.6  $&$0.8  $&$5.5  $&$0.82 $&$6.4$ \\ 
II c&$0.113\,     $&$7.3  $&$17.3 $&$13.3 $&$11.2 $&$8.9  $&$0.8  $&$0.9  $&$6.5  $&$0.84 $&$4.8$ \\ 
II d&$0.00526\,   $&$10.3 $&$24.4 $&$18.7 $&$15.7 $&$13.1 $&$1.1  $&$0.9  $&$8.4  $&$0.87 $&$4.7$ \\ 
\hline 
III a&$0.126\,    $&$5.0  $&$10.8 $&$8.3  $&$6.8  $&$3.7  $&$0.5  $&$0.7  $&$5.5  $&$0.76 $&$10.3$ \\ 
III b&$0.203\,    $&$4.2  $&$10.9 $&$8.2  $&$7.2  $&$3.9  $&$0.7  $&$0.6  $&$5.8  $&$0.81 $&$6.6$ \\ 
III c&$0.0503\,   $&$7.6  $&$13.9 $&$11.1 $&$8.3  $&$4.0  $&$0.8  $&$0.5  $&$7.0  $&$0.86 $&$6.4$ \\ 
III d&$0.00231\,  $&$10.6 $&$22.4 $&$17.5 $&$14.0 $&$11.0 $&$1.1  $&$0.7  $&$8.3  $&$0.87 $&$4.4$ \\ 
\hline 
IV a&$0.0350\,    $&$6.8  $&$13.1 $&$10.4 $&$8.0  $&$5.0  $&$0.7  $&$0.5  $&$5.9  $&$0.78 $&$7.1$ \\ 
IV b&$0.0513\,    $&$5.5  $&$13.9 $&$10.5 $&$9.1  $&$6.6  $&$1.2  $&$0.7  $&$5.9  $&$0.82 $&$7.1$ \\ 
IV c&$0.0167\,    $&$9.7  $&$14.9 $&$12.5 $&$8.1  $&$3.0  $&$1.3  $&$0.8  $&$7.2  $&$0.85 $&$5.3$ \\ 
IV d&$0.000750\,  $&$14.1 $&$20.6 $&$17.6 $&$10.6 $&$7.0  $&$1.4  $&$0.7  $&$7.7  $&$0.85 $&$5.3$ \\ 
\hline 

\end{tabular} 
\normalfont
 \vspace{\baselineskip} 
\caption{\label{tab::Trijet_Q2PT} 
Double differential 3-jet cross sections as 
a function of $Q^2$ and $\Ptav$ of two leading jets
together with their relative errors and 
hadronisation correction 
factors. Other details are given in the caption to table \ref{tab::jet_Q2}. 
The bin nomenclature is defined in table \ref{tab:bins}.} 
\end{table}

\begin{table}[htp]
\centering
\tiny \sf
\renewcommand{\arraystretch}{1.50}
\begin{tabular}{|c || r || r | r | r | r || r | r | r | r || r | r|} 

\multicolumn{12}{c}{ } \\
\multicolumn{12}{c}{\normalsize 3-Jet to 2-Jet Cross Sections Ratio ${\rm 
                    \frac{d\sigma_{3-jet}}{dQ^2}/\frac{d\sigma_{2-jet}}{dQ^2}}$}\\
\multicolumn{12}{c}{ } \\
\hline
 & & &  & total & total &
\multicolumn{4}{|c||}{\underline{\hspace*{0.1cm}single contributions to 
                       correlated uncertainty\hspace*{0.1cm}} }
   &                                           hadronisation & hadronisation\\
bin   & 3-jet/2-jet        & statistical  & total        & uncorrelated & correlated & model  & electron     & electron   & hadronic    & 
correction    & correction    \\
      & ratio      & uncert.      & uncert.      & uncertainty  & uncert.    & uncert.& energy scale & polar angle&  energy scale &  factor    & uncertainty \\
    &              & [\%]    & [\%]         & [\%]         & [\%]       &   [\%] & [\%]         & [\%]       & [\%] &       & [\%]\\
\hline
I   &$0.156\,     $&$2.4  $&$6.8  $&$5.0  $&$4.6  $&$4.2  $&$0.6  $&$0.3  $&$1.2  $&$0.88 $&$4.7$ \\ 
II  &$0.162\,     $&$2.5  $&$6.1  $&$4.6  $&$4.1  $&$3.6  $&$0.7  $&$0.1  $&$1.2  $&$0.88 $&$5.1$ \\ 
III &$0.170\,     $&$2.4  $&$5.1  $&$3.8  $&$3.4  $&$2.6  $&$0.8  $&$0.1  $&$1.4  $&$0.85 $&$6.9$ \\ 
IV  &$0.151\,     $&$3.7  $&$8.0  $&$6.2  $&$5.1  $&$4.6  $&$0.8  $&$0.1  $&$1.5  $&$0.85 $&$6.8$ \\ 
\hline

\multicolumn{12}{c}{ } \\
\multicolumn{12}{c}{\normalsize 3-Jet to 2-Jet Cross Sections Ratio ${\rm 
                    \frac{d\sigma_{3-jet}}{d\Ptav}/\frac{d\sigma_{2-jet}}{d\Ptav}}$}\\
\multicolumn{12}{c}{ } \\
\hline
 & & &  & total & total &
\multicolumn{4}{|c||}{\underline{\hspace*{0.1cm}single contributions to 
                       correlated uncertainty\hspace*{0.1cm}} }
   &                                           hadronisation & hadronisation\\
bin   & 3-jet/2-jet        & statistical  & total        & uncorrelated & correlated & model  & electron     & electron   & hadronic    & 
correction    & correction    \\
      & ratio      & uncert.      & uncert.      & uncertainty  & uncert.    & uncert.& energy scale & polar angle&  energy scale &  factor    & uncertainty \\
    &              & [\%]    & [\%]         & [\%]         & [\%]       &   [\%] & [\%]         & [\%]       & [\%] &       & [\%]\\
\hline
 a&$0.0982\,     $&$2.4  $&$4.1  $&$3.1  $&$2.6  $&$1.2  $&$0.7  $&$0.1  $&$1.5  $&$0.84 $&$8.5$ \\ 
 b&$0.222\,      $&$1.9  $&$3.2  $&$2.4  $&$2.1  $&$1.2  $&$0.4  $&$0.1  $&$0.7  $&$0.86 $&$5.2$ \\ 
 c&$0.276\,      $&$3.5  $&$4.8  $&$4.1  $&$2.5  $&$1.9  $&$0.4  $&$0.1  $&$0.7  $&$0.89 $&$3.0$ \\ 
 d&$0.310\,      $&$5.1  $&$6.8  $&$5.9  $&$3.3  $&$2.9  $&$0.4  $&$0.1  $&$0.7  $&$0.91 $&$2.9$ \\ 
\hline

\multicolumn{12}{c}{ } \\
\multicolumn{12}{c}{\normalsize 3-Jet to 2-Jet Cross Sections Ratio 
                    ${\rm \frac{d^2\sigma_{3-jet}}{dQ^2d\Ptav}/\frac{d^2\sigma_{2-jet}}{dQ^2d\Ptav}}$}\\ 
\multicolumn{12}{c}{ } \\
\hline
 & & &  & total & total &
\multicolumn{4}{|c||}{\underline{\hspace*{0.1cm}single contributions to 
                       correlated uncertainty\hspace*{0.1cm}} }
   &                                           hadronisation & hadronisation\\
bin   & 3-jet/2-jet        & statistical  & total        & uncorrelated & correlated & model  & electron     & electron   & hadronic    & 
correction    & correction    \\
      & ratio      & uncert.      & uncert.      & uncertainty  & uncert.    & uncert.& energy scale & polar angle&  energy scale &  factor    & uncertainty \\
    &              & [\%]    & [\%]         & [\%]         & [\%]       &   [\%] & [\%]         & [\%]       & [\%] &       & [\%]\\
\hline
I a&$0.0980\,     $&$4.1  $&$6.7  $&$5.5  $&$3.9  $&$3.2  $&$0.7  $&$0.2  $&$1.4  $&$0.85 $&$10.2$ \\ 
I b&$0.228\,      $&$3.5  $&$6.6  $&$5.2  $&$4.1  $&$3.7  $&$0.3  $&$0.2  $&$0.7  $&$0.87 $&$5.0$ \\ 
I c&$0.267\,      $&$6.3  $&$12.0 $&$9.5  $&$7.4  $&$7.2  $&$0.3  $&$0.1  $&$0.8  $&$0.87 $&$5.1$ \\ 
I d&$0.294\,      $&$10.2 $&$12.4 $&$11.3 $&$5.2  $&$4.8  $&$0.2  $&$0.3  $&$0.9  $&$0.93 $&$5.3$ \\ 
\hline 
II a&$0.103\,     $&$4.4  $&$6.6  $&$5.5  $&$3.6  $&$2.8  $&$0.8  $&$0.2  $&$1.4  $&$0.84 $&$9.2$ \\ 
II b&$0.215\,     $&$3.7  $&$5.7  $&$4.7  $&$3.3  $&$2.8  $&$0.4  $&$0.1  $&$0.8  $&$0.88 $&$6.0$ \\ 
II c&$0.293\,     $&$6.9  $&$9.7  $&$8.4  $&$5.0  $&$4.7  $&$0.4  $&$0.1  $&$0.5  $&$0.90 $&$4.3$ \\ 
II d&$0.312\,     $&$9.2  $&$12.0 $&$10.7 $&$5.5  $&$5.2  $&$0.5  $&$0.2  $&$0.9  $&$0.91 $&$7.6$ \\ 
\hline 
III a&$0.0966\,   $&$4.6  $&$7.2  $&$6.0  $&$4.1  $&$3.4  $&$0.9  $&$0.2  $&$1.4  $&$0.81 $&$11.0$ \\ 
III b&$0.243\,    $&$3.2  $&$4.9  $&$4.0  $&$2.8  $&$2.2  $&$0.6  $&$0.1  $&$0.9  $&$0.85 $&$6.9$ \\ 
III c&$0.287\,    $&$6.9  $&$9.8  $&$8.4  $&$5.1  $&$4.7  $&$0.4  $&$0.1  $&$1.1  $&$0.90 $&$8.4$ \\ 
III d&$0.359\,    $&$9.4  $&$14.0 $&$11.9 $&$7.4  $&$7.2  $&$0.5  $&$0.1  $&$0.4  $&$0.90 $&$8.5$ \\ 
\hline 
IV a&$0.0922\,    $&$6.7  $&$10.9 $&$9.0  $&$6.2  $&$5.6  $&$0.6  $&$0.2  $&$2.0  $&$0.83 $&$8.8$ \\ 
IV b&$0.193\,     $&$5.5  $&$11.0 $&$8.6  $&$6.8  $&$6.5  $&$0.8  $&$0.1  $&$0.8  $&$0.85 $&$8.8$ \\ 
IV c&$0.257\,     $&$9.0  $&$13.3 $&$11.3 $&$6.9  $&$6.7  $&$0.7  $&$0.2  $&$0.4  $&$0.89 $&$6.6$ \\ 
IV d&$0.282\,     $&$14.1 $&$16.1 $&$15.1 $&$5.6  $&$5.3  $&$0.9  $&$0.2  $&$0.2  $&$0.90 $&$6.3$ \\ 
\hline 

\end{tabular} 
\normalfont
 \vspace{\baselineskip} 
\caption{\label{tab::Ratio_Q2Pt} 
Ratio of 3-jet to 2-jet cross sections as 
a function of $Q^2$ and $\Ptav$ together with their relative errors 
and hadronisation correction 
factors. Other details are given in the caption to table \ref{tab::jet_Q2}. 
The bin nomenclature is defined in table \ref{tab:bins}.} 
\end{table}

\begin{table}[h]
\normalsize
	\centering
	\renewcommand{\arraystretch}{1.80}
   {\large Determination of $\alpha_s$ from jets} \\[3pt]
  \begin{tabular}{| l | c | c | c| c | c |}
			\hline 
     	\multirow{3}{*}{Measurement} & 	\multirow{3}{*}{$\alpha_S(M_Z)$} & \multicolumn{3}{|c|}{Uncertainty} &	\multirow{3}{*}{$\chi^2/\rm ndf$} \\ \cline{3-5} 

     	& & experimental & theory  & PDF & \\

		\hline        
    $\sigma_{\rm jet}\left(Q^2,P_T\right)$ &
    $0.1180$ & $0.0018$ & $^{+0.0122}_{-0.0090}$ & $0.0022$ & $17.5/21$ \\ \hline
    $\sigma_{\textnormal{2-jet}}\left(Q^2,\left\langle P_T \right\rangle \right)$ &
    $0.1155$ & $0.0018$ & $^{+0.0121}_{-0.0090}$ & $0.0025$ & $14.3/23$ \\ \hline
    $\sigma_{\textnormal{3-jet}}\left(Q^2, \left\langle P_T \right\rangle \right)$ &
    $0.1170$ & $0.0017$ & $^{+0.0090}_{-0.0072}$ & 0.$0014$ & $11.0/15$ \\ \hline
\hline
      $\sigma_{\rm jet},~\sigma_{\textnormal{2-jet}},~\sigma_{\textnormal{3-jet}}$ &
    $0.1160$ & $0.0014$ & $^{+0.0093}_{-0.0077}$ & $0.0016$ & $50.6/61$ \\ \hline
        $\sigma_{\textnormal{3-jet}}/\sigma_{\textnormal{2-jet}}$ &
    $0.1215$ & $0.0032$ & $^{+0.0066}_{-0.0058}$ & $0.0013$ & $11.9/13$ \\ \hline 
  \end{tabular}
  \vspace{\baselineskip} 
	\caption{\label{tab::Fits} Values of $\alpha_s (M_Z)$ 
        obtained from fits to the individual inclusive jet, 
        2-jet and 3-jet double differential cross sections
        and from a simultaneous fit to all of 
        them and to the ratio of 3-jet to 2-jet cross sections. 
        Fitted values are given with experimental, 
        theoretical and PDF errors,
        the normalised $\chi^2$/ndf of the fit is also shown.}
\end{table}

\newpage
\clearpage

\vfill


\end{document}